\documentclass[preprint,12pt,3p,fleqn]{elsarticle}
\pdfoutput=1

\usepackage{amssymb}
\usepackage{amsthm}
\usepackage{amsmath}
\usepackage{multirow}
\usepackage{booktabs}
\usepackage{graphicx}
\usepackage{float}
\usepackage{url}
\usepackage{epstopdf}
\usepackage{pdflscape}
\usepackage{amsthm}
\usepackage{amsmath}
\usepackage[english]{babel}
\usepackage{subfigure}
\usepackage{threeparttable}

\usepackage{algorithm}
\usepackage{algorithmic}

\usepackage{hyperref}
\hypersetup{hidelinks,
	colorlinks=true,
	allcolors=black,
	pdfstartview=Fit,
	breaklinks=true}

\usepackage{amsmath}

\usepackage{xcolor}

 \usepackage{booktabs}
 \usepackage{multirow}

\usepackage{tabularx}
\usepackage{graphicx}
\usepackage{caption}

\usepackage{threeparttable}
\usepackage[figuresright]{rotating}
\usepackage[toc,page,title,titletoc,header]{appendix}

\makeatletter

\newcommand{\Rmnum}[1]{\expandafter\@slowromancap\romannumeral #1@}
\makeatother

\journal{XXX}

\begin{document}
	\captionsetup{font={scriptsize}}
	\captionsetup[figure]{labelformat={default},labelsep=period,name={Fig.}}
	
	\begin{frontmatter}
		\title{A new \textcolor{black}{smoothed particle hydrodynamics} method based on high-order moving-least-square \textcolor{black}{targeted essentially non-oscillatory} scheme for compressible flows}
		
		\author[a]{Tianrun Gao}
		\author[a]{Tian Liang}
		\author[a,b,c,d]{Lin Fu\corref{cor1}}
		\ead{linfu@ust.hk}
		\cortext[cor1]{Corresponding author.}

		\address[a]{Department of Mechanical and Aerospace Engineering, The Hong Kong University of Science and Technology, Clear Water Bay, Kowloon, Hong Kong}
		\address[b]{Department of Mathematics, The Hong Kong University of Science and Technology, Clear Water Bay, Kowloon, Hong Kong}
		\address[c]{HKUST Shenzhen-Hong Kong Collaborative Innovation Research Institute, Futian, Shenzhen, China}
		\address[d]{Shenzhen Research Institute, The Hong Kong University of Science and Technology, Shenzhen, China}

		\begin{abstract}

			In this study, we establish a hybrid high-order \textcolor{black}{smoothed particle hydrodynamics (SPH)} framework (MLS-TENO-SPH) for compressible flows with discontinuities, which is able to achieve genuine high-order convergence in smooth regions and also capture discontinuities well in non-smooth regions. The framework can be either fully Lagrangian, Eulerian or realizing arbitary-Lagrangian-Eulerian (ALE) feature enforcing the isotropic particle distribution in specific cases. In the proposed framework, the computational domain is divided into smooth regions and non-smooth regions, and these two regions are determined by a strong scale separation strategy in the \textcolor{black}{targeted essentially non-oscillatory (TENO)} scheme. In smooth regions, the moving-least-square (MLS) approximation is used for evaluating high-order derivative operator, which is able to realize genuine high-order construction; in non-smooth regions, the new TENO scheme based on Vila's framework with several new improvements will be deployed to capture discontinuities and high-wavenumber flow scales with low numerical dissipation. The present MLS-TENO-SPH method is validated with a set of challenging cases based on the Eulerian, Lagrangian or ALE framework. Numerical results demonstrate that the MLS-TENO-SPH method features lower numerical dissipation and higher efficiency than the conventional method, and can restore genuine high-order accuracy in smooth regions. Overall, the proposed framework serves as a new exploration in high-order SPH methods, which are potential for compressible flow simulations with shockwaves.

		\end{abstract}

		\begin{keyword}
			
			Smoothed Particle Hydrodynamics \sep  Hybrid method  \sep  High-order numerical schemes \sep TENO schemes  \sep Shock-capturing schemes \sep Compressible flows
			
			
			
		\end{keyword}
		
	\end{frontmatter}
	
	\section{Introduction}

	\textcolor{black}{SPH is a Lgarangian meshless method, initially proposed by Lucy \cite{lucy1977numerical}, and Gingold and Monaghan \cite{gingold1977smoothed} to simulate astrophysical problems. This method is later introduced to simulate solids \cite{benz1995simulations}\cite{monaghan2000sph} and free-surface flows \cite{monaghan1994simulating}. }
This method has increasingly shown its potential  in a number of scenarios, e.g., free-surface problems \cite{colagrossi2003numerical}\cite{ferrari2009new}\cite{antuono2010free}\cite{violeau2016smoothed}, multiphase problems \cite{colagrossi2003numerical}\cite{wang2016overview}\cite{sun2019study}, fluid-structure interaction (FSI) problems \cite{liu2019smoothed}\cite{sun2021accurate}\cite{zhang2021multi}\cite{gao2022block}\cite{gao2023multi}\cite{gao2023new}, \textcolor{black}{chemistry and biology applications \cite{avesani2017alternative}\cite{herrera2009meshless}\cite{ye2013stretching}}, etc. For comprehensive reviews on the SPH methods, the readers are referred to \cite{monaghan2012smoothed} and \cite{liu2010smoothed}.

	In the classical finite volume (FV) framework, the governing equations are solved using an integral form and discretized with the flux construction based on cell-averaged values. Vila \cite{vila1999particle} proposes an SPH framework similar to FV. The difference is that SPH is a nonlocal method while FV is local, and this feature has been indicated in many pioneering works \cite{hietel2000finite}\cite{vila1999particle}\cite{junk2003finite}. Among these, Vila's SPH formulations \cite{vila1999particle} are used widely due to its rigorousness and simplicity. The Riemann solver is employed for evaluating the interaction between particle pairs, which replaces the function of artificial viscosity in traditional SPH methods. This framework is shown to capture discontinuities very well, similar to FV. While traditional SPH methods are devised for incompressible flows, with this FV-like framework, it is possible to simulate compressible flows with discontinuities or high-wavenumber under-resolved flow scales. 
	
	Though SPH methods have seen tremendous applications in these fields and similarities with FV, the low convergence order is usually a critical problem in SPH. The traditional discretization of SPH is usually around zero-th order \cite{dilts1999moving}\cite{monaghan2005smoothed} when the particle distribution is in disorder. With the Lagrangian feature, the smoothed kernel function renders the simulation stable through relating to many neighboring particles. However, this also leads to inconsistency problem. Therefore, the discretized forms for high-order operators in traditional SPH methods are usually inaccurate, even if all the primitive field values for these operator constructions are accurate. 
	
	There are several pioneering works striving to improve the accuracy of SPH. Based on the SPH discretization, many studies try to construct high-order field values. Several studies transform traditional SPH into Riemann-based formulations, and the interactions between particle pairs are solved with high-order constructions along two interactive particles \cite{zhang2019weakly}\cite{wang2021new}\cite{meng2022targeted}. However, this construction is not truly high-order since a linear interpolation is used to construct the field values of missing particles. \textcolor{black}{For the high-order construction of field variables in two or three-dimensional space, the MLS interpolations are proposed \cite{li1998synchronized}\cite{li1999reproducing0}, and are extended to simulations of fluid mechanics \cite{li1999reproducing1}. This meshless interpolation method is later incorporated into the updated Lagrangian particle hydrodynamics to simulate Newtonian flows \cite{tu2017updated} and multiphase flows \cite{yan2019updated}, where the corresponding high-order methods \cite{yan2020higher} are devised based on the MLS interpolation and the convergence property is proved \cite{yu2021approximation}.}
    For the high-order reconstruction of numerical fluxes,
	Vila \cite{vila1999particle} incorporates the linear MUSCL scheme in the new framework. Following Vila's framework, there are several even higher-order construction methods to simulate compressible flows. Weighted essentially non-oscillatory (WENO) reconstruction methods \cite{jiang1996efficient} are widely employed for high-order construction of field values. Avesani et al. \cite{avesani2014new} propose a WENO reconstruction based on the MLS  method, where nine stencils are deployed for the MLS construction. An improved version is reported in \cite{avesani2021alternative}, several similar studies are also shown in \cite{rossi2017well}\cite{antona2021towards}. Apart from WENO reconstruction, the so-called multidimensional optimal order detection (MOOD) method is also implemented into Vila's framework for stability and low numerical dissipation \cite{nogueira2016high}. There is also a set of other high-order Riemann-based frameworks to increase the SPH accuracy order \cite{parshikov2000improvements}\cite{inutsuka2002reformulation}\cite{parshikov2002smoothed}\cite{cha2003implementations}. 
	
	Indeed, it is noteworthy that even if the reconstructed field values are accurate, the high-order convergence is not achieved in many scenarios for SPH, due to the low accuracy of SPH discretization operators like the gradients, especially when the particle distribution is in disorder. Several studies strive to resolve this issue by constructing high-order operators in SPH. A WENO-MLS method is proposed in \cite{vergnaud2023investigations}, and a theoretical analysis is given for the gradient operator for the one-dimensional case, showing that the high-order accuracy can be achieved for the equally-spaced particles in the Eulerian framework. However, it may be difficult to have the latticed particle distribution within the SPH simulation process. In the work of Nasar et al. \cite{nasar2021high}, a Taylor-series expansion analysis is utilized, and by combining with the SPH discretization, a general modification matrix is derived to increase the discretization accuracy of the gradient operator to the 4th order to simulate the SPH problems in the Eulerian framework. In this method, the Bonet and Lok’s correction \cite{bonet1999variational} is a special case. However, this method may fail when there exist non-smooth discontinuities and high-wavenumber under-resolved flow scales, since the method is based on the Taylor-series expansion. Another way for high-order operator construction is based on the MLS method. In the works of Ramírez et al. \cite{ramirez2018very}\cite{ramirez2022arbitrary}, the gradient operator is constructed using a MLS construction, based on which, a new form of kernel function is proposed. Similarly, this method can achieve the high-order accuracy when the flow field is smooth, however, for discontinuities it may not be applicable since MLS will be highly oscillatory when the flow field is non-smooth. A similar idea has also been developed in \cite{samulyak2018lagrangian}, where a Lagrangian particle method is proposed based on the so-called generalized finite difference (GFD) method \cite{benito2001influence}, and the high-order derivative is obtained with optimal MLS constructions. In these high-order methods, the exact conservation property is usually not guaranteed, as numerical fluxes cannot cancel each other in the system when a high-order modification term is added. Actually, albeit the drawback of non-exact conservation, the high-order convergence of the resultant solution is achieved, and therefore the conservation property should not be a big issue for high-order SPH methods since the numerical errors will vanish eventually when refining the particle resolution. 

\textcolor{black}{In the present study, a hybrid high-order SPH framework for compressible flows with discontinuities will be established. The main contributions of the present study are listed as follows: (a) in the proposed framework, the computational domain is divided into smooth regions and non-smooth regions, and these two regions are determined by a strong scale separation strategy in the TENO scheme \cite{fu2016family}\cite{fu2019improved}\cite{fu2019low}\cite{fu2019hybrid}\cite{fu2021very}\cite{takagi2022novel}\cite{liang2022fifth}\cite{ji2022class}\cite{fu2023review}; (b) in smooth regions, the MLS approximation is used to evaluate the high-order derivative operator, which is able to realize genuine high-order construction, and in non-smooth regions, the new TENO scheme based on Vila's framework with several new improvements will be deployed to capture discontinuities and high-wavenumber flow scales with low numerical dissipation; (c) numerical results demonstrate that the proposed method features lower numerical dissipation and higher efficiency than the conventional SPH method, and can restore genuine high-order accuracy in smooth regions. }

     This paper is organized as follows. In section 2, the Vila's SPH formulations based on the ALE framework will be introduced, where the corresponding Riemann solver is also used. In section 3, an improved high-order TENO scheme for Vila's ALE SPH formulations will be proposed, where the MLS method is used for high-order construction, and compact candidate stencils for different orders will be deployed. Based on the TENO nonlinear weighting strategy, the smooth and non-smooth flow scales will be separated. Then the MLS method is used for evaluating the high-order derivative in the smooth regions, and Vila's framework is deployed in the non-smooth regions. In section 4, a set of challenging cases in the Eulerian, Lagrangian or ALE framework are simulated with the new high-order SPH framework. The linear advection case demonstrates that when the flow field is smooth, the present framework can restore a genuine high-order accuracy. Other simulations of compressible flows show that the present method features lower numerical dissipation and higher computational efficiency when compared to previous state-of-the-arts.

	\section{SPH formulations}
	\subsection{Governing equations for compressible flows}

	The conservative form of Euler equations in the Eulerian framework can be  written as
	\begin{equation}
	\label{eq:ns_c}
	\dfrac{\partial \mathbf{U}}{\partial t} + \nabla \cdot \mathbf{F(\mathbf{U})} = 0,
	\end{equation}
	where $ \mathbf{U} $ is the conservative variable vector and $ \mathbf{F(\mathbf{U})}=(\mathbf{f}, \mathbf{g}) $ is the flux tensor.   $ \mathbf{U} $ and $ \mathbf{F(\mathbf{U})} $ are expressed as
	\begin{equation}
	\mathbf{U}=
	\begin{pmatrix}
	\rho \\
	\rho u\\
	\rho v\\
	E
	\end{pmatrix},
	\quad
	\mathbf{F(\mathbf{U})}=
	\begin{pmatrix}
	\rho u & \rho v\\
	\rho u^2+p & \rho uv\\
	\rho uv & \rho v^2+p\\
	u(E+p) & v(E+p)
	\end{pmatrix}.
	\end{equation}
 	Herein, $\rho$, $p $, $ \boldsymbol{v}(u, v) $ and $ e $ denote density, pressure, velocity and internal energy per unit mass, where $ u $ and $ v $ are two velocity components in the $ x $ and $ y $ directions, respectively. \textcolor{black}{For the calorically ideal gases}, the equation of state (EOS) of $ e $ is expressed as \textcolor{black}{\cite{toro2013riemann}\cite{avesani2021alternative}\cite{ji2019new}}
	\begin{equation}
	e=\dfrac{p}{(\gamma-1)\rho},
	\end{equation}
	where $ \gamma $ is the  ratio of specific heat capacities. $ c $ is the sound speed written as $ \sqrt{\gamma \frac{p}{\rho}} $, and total energy density $ E $ is 
	\begin{equation}
	E=\rho e +\frac{1}{2} \rho (u^2+v^2), 
	\end{equation}
	and therefore  pressure $ p $ is 
	\begin{equation}
	p=(\gamma-1)\left[ E-\frac{1}{2}\rho (u^2+v^2)\right].
	\end{equation}
	The above Eulerian framework can be generalized to an ALE framework. \textcolor{black}{In the ALE framework, the transport derivative $\frac{\tilde{d}}{dt} $ is employed instead of the material derivative $\frac{{d}}{dt} $ in the Lagrangian framework, with the latter as a special case of the former}. If we define transport derivative of the conservative variable vector $ \frac{\tilde{d}\mathbf{U}}{dt} $ as
	\begin{equation}
	\dfrac{\tilde{d}\mathbf{U}}{dt}=\dfrac{\partial\mathbf{U}}{\partial t}+ \nabla \cdot (  \mathbf{U} \otimes \boldsymbol{w} ),
	\label{ale}
	\end{equation}
	where $ \boldsymbol{w} $ is the so-called transport velocity. Then Eq. (\ref{eq:ns_c}) can be transformed as
	\begin{equation}
	\label{eq:ns_ale}
	\dfrac{\tilde{d}\mathbf{U}}{dt} + \nabla \cdot \mathbf{H(\mathbf{U})}     = 0,
	\end{equation}
	where the flux is $ \mathbf{H(\mathbf{U}, \boldsymbol{w} )}=\mathbf{F(\mathbf{U})} -  \mathbf{U} \otimes \boldsymbol{w}$. Specifically, when $ \boldsymbol{w}= \boldsymbol{v} $, the ALE form presents the total Lagrangian form; when  $ \boldsymbol{w}= \boldsymbol{0} $, the ALE form becomes Eulerian form. 
\textcolor{black}{In the ALE framework, the transport velocity $\boldsymbol{w}$ is not directly relevant to the fluid physical variables, but usually $\boldsymbol{w}$ is used to render particle distribution isotropic in the SPH method. In this study, isotropic particle distribution is required to improve the numerical stability, and therefore $\boldsymbol{w}$ is used for isotropic particle distribution in specific cases. The calculation of $\boldsymbol{w}$ is referred to Eq. (\ref{eq:background_P}).}

	\subsection{SPH formulations of Vila}
	Vila \cite{vila1999particle} deduces a set of SPH formulations based on the conservative variable, which is similar to the FV framework. In this study, the fourth-order Wendland kernel \cite{wendland1995piecewise} is employed with a smoothing length of $h=2\Delta {x}$, where $\Delta {x}$ is the initial particle spacing. Referring to the modified work \cite{avesani2014new}, the Lagrangian form is written as
	\begin{equation}
	\label{eq:vila}
	\left\{
	\begin{array}{l}
	\begin{aligned}
	\vspace{1ex}
	&\dfrac{d V_i \mathbf{U}_i}{d t}=-\sum_j^N  2 (\mathbf{G}_{i j}-\mathbf{H}_i) \cdot \nabla W_{i j}V_i V_j,  \\
	&\dfrac{d V_i}{d t}=\sum_j^N 2\left(\overline{\boldsymbol{v}}_{i j}-\boldsymbol{v}_i\right) \cdot \nabla W_{i j} V_i V_j, \\
	&\dfrac{d \boldsymbol{r}_i}{d t}=\boldsymbol{v}_i,
	\end{aligned}
	\end{array}
	\right.
	\end{equation}
	where $\mathbf{G}_{i j}  $ is the numerical flux between particle $ i $ and $ j $, and \textcolor{black}{like FV framework,  the numerical flux is constructed with a Riemann solver in this study}. 
	$  V_i $ denotes the volume of the particle $ i $, and the volume evolution is governed by the second equation in Eq. (\ref{eq:vila}), where the velocity flux $ \overline{\boldsymbol{v}}_{i j}-\boldsymbol{v}_i $ controls the volume evolution, with $ \overline{\boldsymbol{v}}_{i j}=\frac{1}{2}(\boldsymbol{v}_i+\boldsymbol{v}_j) $. 
	Due to the varying particle volume, the smoothing length $ h $ is also varying, and updated based on
	\begin{equation}
	h_i=2\sqrt[v]{V_i},
	\end{equation}
    where $ v $ is the number of dimensions, and $ v=2 $ in 2D. 
	The third equation in Eq. (\ref{eq:vila}) shows that the particle moves with velocity $ \boldsymbol{v}_i $. The difference between Vila's SPH framework and the FV framework is that the flux integration is implemented within kernel support instead of using  Gaussian integrations on the cell interface in the  FV framework. The FV framework is usually Eulerian, and therefore the volume is invariant. Although there are several attempts \cite{barlow2016arbitrary}\cite{barlow2014constrained}\cite{sambasivan2013finite} to extend FV method to Lagrangian framework, it is non-trivial to handle the mesh topology changes and the potential stability issues in these works. By contrast, the discretized kernel form in SPH methods evades many mesh reconstructions, and using the kernel support to construct the relationship with neighbors can ensure the numerical stability.

    In many simulations, the particle distribution is not isotropic in the Lagrangian framework, and therefore the ALE framework is needed to regularize the particle distribution during the particle evolution. Combining Eq. (\ref{ale}) and (\ref{eq:vila}), the discretized form for ALE framework is
	\begin{equation}
	\label{eq:ale_vila}
	\left\{
	\begin{array}{l}
	\begin{aligned}
	\vspace{1ex}
	&\dfrac{\tilde{d} V_i \mathbf{U}_i}{d t}=-\sum_j^N  2 (\tilde{\mathbf{G}}_{i j}-\mathbf{H}_i) \cdot \nabla W_{i j}V_i V_j,  \\
	&\dfrac{\tilde{d} V_i}{d t}=\sum_j^N  2\left(\overline{\boldsymbol{w}}_{i j}-\boldsymbol{w}_i\right)  \cdot \nabla W_{i j} V_i V_j, \\
	&\dfrac{\tilde{d} \boldsymbol{r}_i}{d t}=\boldsymbol{w}_i,
	\end{aligned}
	\end{array}
	\right.
	\end{equation}
	where  $\tilde{\mathbf{G}}_{i j}  $ is the numerical flux between particle $ i $ and $ j $ in the ALE framework, and $ \overline{\boldsymbol{w}}_{i j}=\frac{1}{2}(\boldsymbol{w}_i+\boldsymbol{w}_j) $. 
	
	The transport velocity $ \boldsymbol{w} $ is an additional variable irrelevant to the governing Euler equation itself, and it can be defined freely. In particle methods, the particle distribution has a very important effect on the numerical stability, and an isotropic particle distribution is desired in flow simulations, especially for incompressible flows. Herein, the transport velocity $ \boldsymbol{w} $ is also used to render particle distribution isotropic in the flow field. Assuming $ \boldsymbol{w} = \boldsymbol{v}+\delta \boldsymbol{v} $, then, based on the widely-used Fick's law \cite{xu2009accuracy}, $ \delta \boldsymbol{v} $ is expressed as
	\begin{equation}
	\label{eq:background_P}
	\left\{
	\begin{array}{l}
	\begin{aligned}
	&\delta \boldsymbol{v}^{\prime} = - 0.25ch \sum_{j} \left[1+0.2\left(\dfrac{W_{i j}}{W(\Delta x,h_i)}\right)^{4}  \right] {\nabla} W_{i j} V_{j},\\
	&\delta \boldsymbol{v} = \min\left(0.5\left| \boldsymbol{v}\right|, \left| \delta \boldsymbol{v}^{\prime} \right| \right)\dfrac{\delta \boldsymbol{v}^{\prime}}{\left| \delta \boldsymbol{v}^{\prime} \right|},
	\end{aligned}
	\end{array}
	\right.
	\end{equation}
	where $ c $ is the sound speed of the fluid. The second equation ensures that $ \delta \boldsymbol{v} $ is not too large, which may affect the numerical accuracy.

	In the work of Nasar et al. \cite{nasar2021high}, a Taylor-series expansion is used for reconstructing the flow field value, and combining with SPH discretization, a modification matrix is derived to increase the discretization accuracy of the gradient operator to fourth order. In the present study, a second-order modification matrix $ \mathbf{M}_{i} $ is used to increase the local accuracy of the gradient operator, then the governing equations are written as
	\begin{equation}
	\mathbf{M}_{i} =
	\begin{pmatrix}
	\sum_j x_{i j} W_{i j}^x V_j & \sum_j y_{i j} W_{i j}^x V_j \\
	\sum_j x_{i j} W_{i j}^y V_j & \sum_j y_{i j} W_{i j}^y V_j
	\end{pmatrix},
	\end{equation}
	where $  W_{i j}^x $ and $  W_{i j}^y $ are two components in kernel gradient $\nabla W_{i j}=(W_{i j}^x, W_{i j}^y)^T$.  Then, based on Eq. (\ref{eq:ale_vila}), we have
	\begin{equation}
	\label{eq:ale_vila1}
	\left\{
	\begin{array}{l}
	\begin{aligned}
	\vspace{1ex}
	&\dfrac{\tilde{d} V_i \mathbf{U}_i}{d t}=-\sum_j^N  2 (\tilde{\mathbf{G}}_{i j}-\mathbf{H}_i) \cdot \mathbf{M}_{i}^{-1}\nabla W_{i j}V_i V_j,  \\
	&\dfrac{\tilde{d} V_i}{d t}=\sum_j^N  2\left(\overline{\boldsymbol{w}}_{i j}-\boldsymbol{w}_i\right)  \cdot \mathbf{M}_{i}^{-1}\nabla W_{i j} V_i V_j. \\
	&\dfrac{\tilde{d} \boldsymbol{r}_i}{d t}=\boldsymbol{w}_i,
	\end{aligned}
	\end{array}
	\right.
	\end{equation}
	This formulation leads to a second-order discretization accuracy when the flow field is smooth. 
	
	\subsection{Numerical fluxes of particle interaction}
	\label{sect:riemann}
	The proper design of numerical flux is crucial in compressible flows, especially when discontinuity exists. In this study, the Osher-type \cite{dumbser2011universal} Riemann solver is used due to its low-dissipation nature. For $ \mathbf{H(\mathbf{U}, \boldsymbol{w} )}=\mathbf{F(\mathbf{U})} - \mathbf{U} \otimes\boldsymbol{w} $, the Osher-type numerical flux \cite{dumbser2011universal} can be expressed as
	\begin{equation}
    \label{eq:Riemann_flux}
	\tilde{\mathbf{G}}_{i j}\left(\mathbf{U}_i^{-}, \mathbf{U}_j^{+}, \overline{\boldsymbol{w}}_{i j} \right)=\frac{1}{2}\left[\mathbf{H}\left(\mathbf{U}_j^{+}, \overline{\boldsymbol{w}}_{i j}\right)+\mathbf{H}\left(\mathbf{U}_i^{-}, \overline{\boldsymbol{w}}_{i j}\right)\right]-\boldsymbol{\Theta}\left(\mathbf{U}_j^{+}-\mathbf{U}_i^{-}\right) \otimes \boldsymbol{n}_{i j},
	\end{equation}
	where $ \boldsymbol{n}_{i j} $ is the unit vector pointing from particle $ i $ to $ j $. $ \mathbf{U}_i^{-} $ and $ \mathbf{U}_j^{+} $ denote the two reconstructed conservative variables at the interface between particles $ i $ and $ j $, which will be introduced in section \ref{sect:teno}, and the reconstructions for the variables are referred to Eq. (\ref{eq:comb_s1}) or (\ref{eq:comb_s2}). The Jacobian matrix $ \mathbf{A}_n $ of the flux $\mathbf{H(\mathbf{U}, \boldsymbol{w} )} $ calculated in the direction of $ \boldsymbol{n}_{i j} $ is 
	\begin{equation}
	\mathbf{A}_n= \dfrac{\partial \mathbf{H}}{\partial \mathbf{U}} \cdot \boldsymbol{n}_{i j}= (\dfrac{\partial \mathbf{f}}{\partial \mathbf{U}} - \omega_1\mathbf{I}, \dfrac{\partial \mathbf{g}}{\partial \mathbf{U}} - \omega_2\mathbf{I})\cdot \boldsymbol{n}_{i j},
	\end{equation}
    where $\omega_1$ and $\omega_2$ are the vector components of $\boldsymbol{w}$ in $x$ and $y$ directions, respectively. Herein, $\frac{\partial \mathbf{f}}{\partial \mathbf{U}}$ and $\frac{\partial \mathbf{g}}{\partial \mathbf{U}}$ are expressed as
	\begin{equation}
	\dfrac{\partial \mathbf{f}}{\partial \mathbf{U}}=\mathbf{A}=
	\begin{pmatrix}
	0 & 1 & 0 & 0\\
	\frac{1}{2}(\gamma-1)q^2-u^2 & (3-\gamma)u & -(\gamma-1)v & \gamma-1 \\
	-uv & v& u & 0\\
    [\frac{1}{2}(\gamma-1)q^2- \frac{E+P}{\rho}]u & \frac{E+P}{\rho} -(\gamma-1)u^2& -(\gamma-1)uv & \gamma u
	\end{pmatrix},
	\end{equation}
and
	\begin{equation}
	\dfrac{\partial \mathbf{g}}{\partial \mathbf{U}}=\mathbf{B}=
	\begin{pmatrix}
	0 & 0 & 1 & 0\\
    -uv & v& u & 0\\
	\frac{1}{2}(\gamma-1)q^2-v^2 & -(\gamma-1)u & (3-\gamma)v & \gamma-1 \\
    [\frac{1}{2}(\gamma-1)q^2- \frac{E+P}{\rho}]v &  -(\gamma-1)uv &\frac{E+P}{\rho}-(\gamma-1)u^2& \gamma v
	\end{pmatrix},
	\end{equation}
 where $q^2 = u^2 +v^2$. Then $ \boldsymbol{\Theta} $ is defined as
	\begin{equation}
	\label{eq:osher}
	\boldsymbol{\Theta}=\frac{1}{2} \int_{0}^{1} \left| \mathbf{A}_n\left[ \mathbf{\Xi}(s), \overline{\boldsymbol{w}}_{i j}\right]\right|  ds.
	\end{equation}
	Here, $ \left| \mathbf{A}_n\right|  $ is written as $ \left| \mathbf{A}_n\right| = \mathbf{V} \left| \mathbf{\Lambda}\right| \mathbf{V}^{-1} $, where $ \mathbf{V} $ is the eigenvector of $  \mathbf{A}_n $, and $ \mathbf{\Lambda} $ is the eigenvalue matrix. 
	$ \mathbf{\Xi}(s) $ is written as
	\begin{equation}
	\mathbf{\Xi}(s)=\mathbf{U}_i^{-} + s(\mathbf{U}_j^{+}-\mathbf{U}_i^{-}).
	\end{equation}
	In practice, Eq. (\ref{eq:osher}) can be calculated by an approximate Gaussian quadrature method, and three-point approximation is used in this study.

    \section{High-order SPH scheme}
    In this section, a high-order SPH scheme will be developed. Firstly, the general MLS method is employed for the high-order data reconstruction for any given stencil. Secondly, the high-order nonlinear TENO reconstruction scheme, featuring lower numerical dissipation, for non-smooth flows based on the SPH framework is proposed. Thirdly, the high-order approximation for gradient operator in smooth regions is also developed based on the coefficients of the  MLS data construction in the corresponding stencils. Finally, a unified SPH framework which can restore high-order numerical accuracy in the smooth flows and capture discontinuities in the non-smooth flows is devised. 
	
	\subsection{MLS data reconstruction method}
	\label{sect:mls}
	MLS is a widely used method for high-order data reconstruction in meshless methods. It has been employed to increase the accuracy in several incompressible simulations with SPH methods \cite{colagrossi2003numerical}\cite{marrone2011delta}. In 2D, for a field value $ \tilde{u}(\tilde{x},\tilde{y}) $, it can be approximated as
	\begin{equation}
	\label{eq:msl1}
	\tilde{u}(\tilde{x},\tilde{y}) = u_i(x_i,y_i) + \boldsymbol{N}^T \mathbf{D} \boldsymbol{b}=u_i(x_i,y_i) + \boldsymbol{N}^T  \boldsymbol{b}^\prime,
	\end{equation}
	where $ u_i(x_i,y_i) $ is the value at particle $ i $. The non-dimensionalized shape function vector $ \boldsymbol{N} $ with $ m $ elements is written as
	\begin{equation}
	\boldsymbol{N} = \left[\xi, \eta,\xi^2, \eta^2, \xi \eta, \xi^3, \eta^3,  \xi^2 \eta, \xi \eta^2, \xi^4, \eta^4,  \xi^3 \eta, \xi \eta^3, \xi^2 \eta^2, \ldots \right]^T,
	\end{equation}
	and the diagonal matrix $ \mathbf{D} $ is
	\begin{equation}
	\mathbf{D} = diag \left(h,h,h^2,h^2,h^2,h^3,h^3,h^3,h^3,h^4,h^4,h^4,h^4,h^4,\ldots \right),
	\end{equation}
	where $ \xi=(\tilde{x}-x_i)/h $ and $ \eta=(\tilde{y}-y_i)/h $. $ \boldsymbol{b} $ is the corresponding coefficient vector. 
	Given a set of particles, $ \boldsymbol{b} $  can be obtained using the least-square principle, as
	\begin{equation}
	\label{eq:msl2}
	\boldsymbol{b} = \mathbf{D}^{-1} \left[ \sum_{j} \boldsymbol{N}_j\boldsymbol{N}^T_j w(\left| \boldsymbol{r}_{ij} \right|) V_j\right] ^{-1} \sum_{j} (u_j - u_i)\boldsymbol{N}_j  w(\left| \boldsymbol{r}_{ij} \right|) V_j,
	\end{equation}
	then the field variable can be reconstructed near particle $ i $. The weighting function $  w(\left| \boldsymbol{r}_{ij} \right| ) $ is referred to \cite{nogueira2016high} as
	\begin{equation}
	\label{eq:msl3}
	w(\left| \boldsymbol{r}_{ij} \right| ) = \dfrac{e^{- {\left( \frac{\left| \boldsymbol{r}_{ij} \right|}{r_m}\right) }^2} -e^{-1} }{1-e^{-1}},
	\end{equation}
	where $ {r_m} = 2\max(\left| \boldsymbol{r}_{ij} \right|) $.
	
	In 2D, there are 9 elements in the shape function vector for the fourth-order approximation, and 14 elements for fifth-order. Usually, when the number of particles $ K $ is needed for a $ n $th-order approximation, there should be at least twice of the element number in the shape function vector \cite{tsoutsanis2014weno}\cite{tsoutsanis2019stencil}, i.e., $ K>n(n+1)-2 $, otherwise the insufficient number of particles may lead to the ill condition of the matrix. 
	
	\subsection{High-order TENO reconstruction scheme for non-smooth flows}
	\label{sect:teno}
	In the highly compressible flows, there are usually discontinuities, and the deployment of a high-order reconstruction scheme crossing the discontinuity will lead to oscillatory solutions. For avoiding such artificial numerical oscillations, a set of candidate stencils with nonlinear combinations can be deployed to reconstruct the field variable at the interface of particle interactions. The WENO scheme \cite{jiang1996efficient} is the most widely used reconstruction method for shock capturing, which ensures  non-oscillatory solutions. In this study, the TENO scheme proposed by Fu et al. \cite{fu2016family}\cite{fu2019low}\cite{ji2022class} is employed for the high-order reconstruction, since TENO features lower numerical dissipation than WENO scheme of the same accuracy order. 
	
\begin{figure}[htbp]
		\centering
		\includegraphics[width=0.6\textwidth]{ 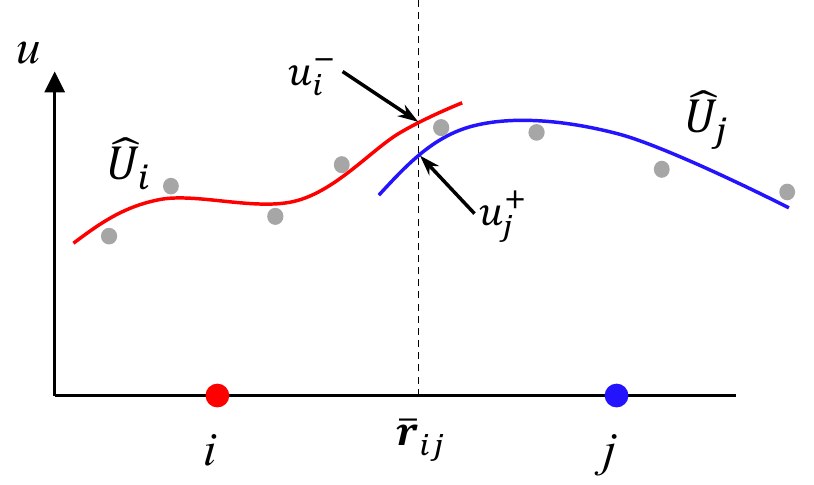}
		\caption{The schematic of the left and right states: $ \hat{U}_i $ and $ \hat{U}_j $ are constructed variables using Eq. (\ref{eq:comb_s1}) or (\ref{eq:comb_s2}), for particles $ i $ and $ j $. The left state $ u_i^- $ and right state $ u_j^+ $ at the interface $ \overline{\boldsymbol{r}}_{ij} $ are derived by $ u_i^-=\hat{U}_i(\overline{\boldsymbol{r}}_{ij}) $ and $ u_j^+=\hat{U}_j(\overline{\boldsymbol{r}}_{ij}) $, respectively. }
		\label{U_LR}
	\end{figure}

	The objective of the reconstruction scheme is to reconstruct the field variables (left state $ \mathbf{U}_i^- $ and right state $ \mathbf{U}_j^+ $) at the interface of two particles $ i $ and $ j $. With the obtained left and right states, the fluxes can be estimated using a Riemann solver, as introduced in section \ref{sect:riemann}. 
	Fig. \ref{U_LR} illustrates the schematic of the left and right states between particles $ i $ and $ j $. Firstly, the high-order constructions of field variable $ \hat{U}_i $ and $ \hat{U}_j $ for particles $ i $ and $ j $ can be obtained using the reconstructed variables, i.e., Eq. (\ref{eq:comb_s1}) or (\ref{eq:comb_s2}). Then the left state $ u_i^- $ and right state $ u_j^+ $ at the interface $ \overline{\boldsymbol{r}}_{ij} $ are derived by calculating $ u_i^-=\hat{U}_i(\overline{\boldsymbol{r}}_{ij}) $ and $ u_j^+=\hat{U}_j(\overline{\boldsymbol{r}}_{ij}) $. Here, the interface position $ \overline{\boldsymbol{r}}_{ij} $ is expressed as
	\begin{equation}
	\overline{\boldsymbol{r}}_{ij} = \dfrac{h_j \boldsymbol{r}_i + h_i \boldsymbol{r}_j }{h_i+h_j},
	\end{equation}
	where $ h_i $ and $ h_j $ denote the smoothing length of particles $ i $ and $ j $, respectively.

	\subsubsection {Candidate stencil arrangement}
	
\begin{figure}[htbp]
		\centering
		\includegraphics[width=0.9\textwidth]{ 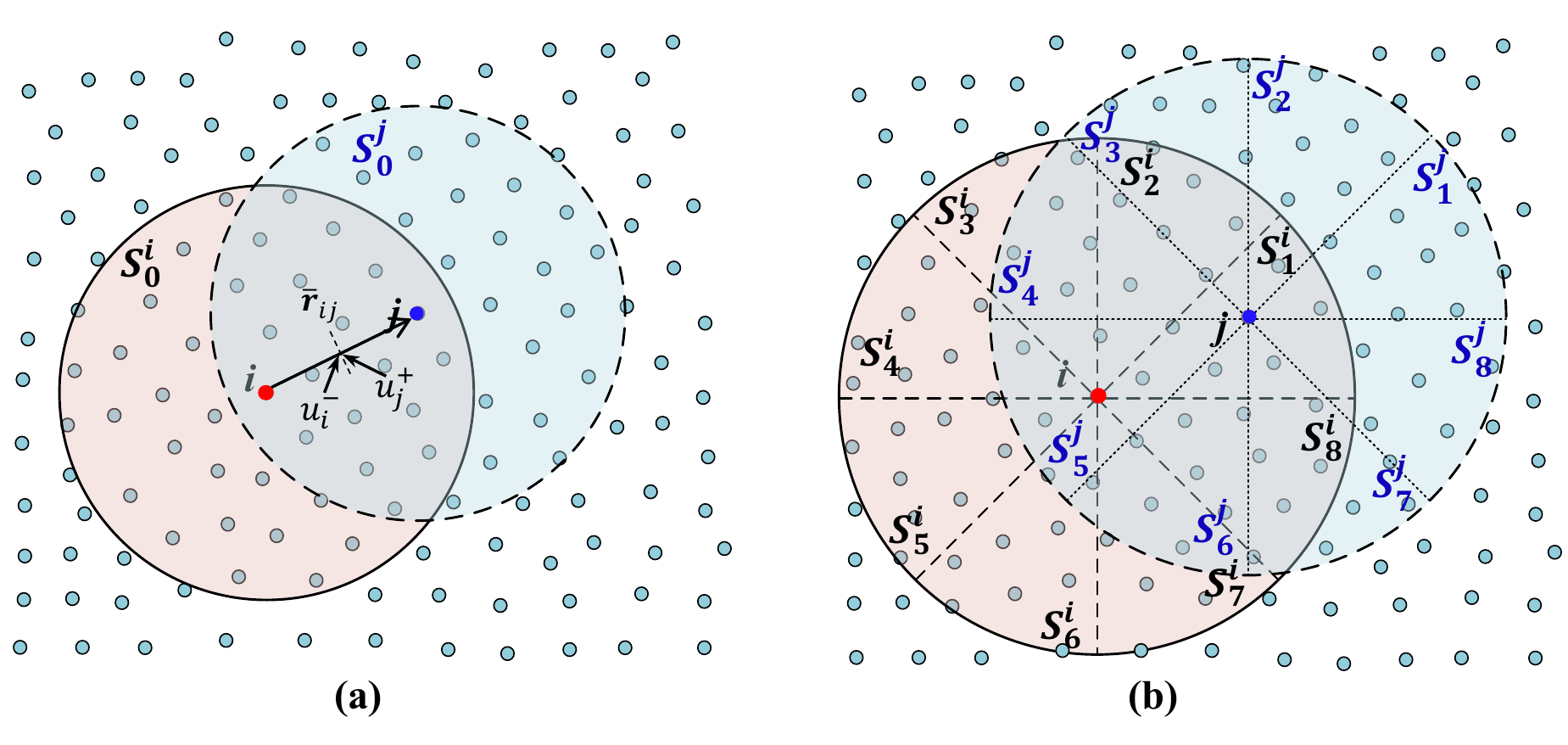}
		\caption{The schematic of (a) the large stencil $ \mathcal{S}^i_0 $ and $ \mathcal{S}^j_0 $ and (b) small directional stencils $ \mathcal{S}^i_1  \sim  \mathcal{S}^i_8 $ and $ \mathcal{S}^j_1  \sim  \mathcal{S}^j_8 $ for particle $i$ and particle $j$, respectively. The stencils are employed for the reconstructions of the left state $ u_i^- $ and right state $ u_j^+ $ (also illustrated in Fig. \ref{U_LR}) at the interface $ \overline{\boldsymbol{r}}_{ij} $, which is used to calculate the Riemann solver in Eq. (\ref{eq:Riemann_flux}). }
		\label{sten}
	\end{figure}

	In the TENO scheme, the candidate stencil arrangement is critical to ensure the non-oscillatory property. The deployment of the candidate stencils is shown in Fig. \ref{sten}, where one large central stencil and eight small directional stencils are used. For the large central stencil, a higher-order MLS approximation, i.e., fourth-, fifth-, or sixth-order, can be deployed, which is supposed to be used when the solution is smooth; for the small directional stencils, the third-order MLS approximation is deployed. When the large central stencil is not smooth, the combination of the smooth ones among these small stencils will be adopted for the final reconstruction. 
	
	The particle selection for a given candidate stencil is critical to ensure a robust MLS approximation of a field value. For the fourth-order MLS approximation on the large central stencil, there should be at least 18 particles participating in the reconstruction for robustness, and the radius of the central stencil $ R_i^{(0)} $ for particle $ i $ is chosen to be $ R_i^{(0)}=2.5\sqrt{V_i} $. For the fifth-order MLS approximation on the large central stencil, there should be at least 28 particles to ensure robustness of MLS, according to section \ref{sect:mls}, and in this case, the radius of the central stencil $ R_i^{(0)} $ for particle $ i $ is chosen to be $ R_i^{(0)}=3.2\sqrt{V_i} $. For the sixth-order MLS approximation on the large central stencil, there should be at least 40 particles to ensure the robustness of MLS, according to section \ref{sect:mls}, and in this case, the radius of the central stencil $ R_i^{(0)} $ for particle $ i $ is chosen to be $ R_i^{(0)}=4\sqrt{V_i} $.

	For the third-order MLS approximation on the small directional  stencils, there should be at least 10 particles in each stencil, and the radius of the $ s $th directional stencil $ R_i^{(s)} $ is chosen to be $ R_i^{(s)}=4.5\sqrt{V_i} $. 
	Usually, when the large central stencil is not used, it means that there may be discontinuity in the central stencil. Under this circumstance, a high-order reconstruction on the small directional stencils is not meaningful, since the numerical method is normally first-order accurate near the discontinuity. Therefore, a third-order reconstruction is deployed for the small directional stencils in this work. 
	
	To select particles for the central stencil of particle $ i $, the neighboring particle $ j $ will be selected: $ j\in S_0 $, if $ \left| \boldsymbol{r}_{ij} \right| < 2.5\max(\sqrt{V_i}, \sqrt{V_j}) $ for the fourth-order reconstruction (O4); or if $ \left| \boldsymbol{r}_{ij} \right| < 3.2 \max(\sqrt{V_i}, \sqrt{V_j}) $ for fifth-order (O5); or if $ \left| \boldsymbol{r}_{ij} \right| < 4 \max(\sqrt{V_i}, \sqrt{V_j}) $ for sixth-order (O6). For the 8 small directional stencils, the neighboring particle $ j $ will be selected in $ s $-th stencil: $ j\in S_s $, if $ \pi (s-1)/4 \le \tan^{-1}(\frac{y_{ij}}{x_{ij}}) \le \pi s/4$ and $  \left| \boldsymbol{r}_{ij} \right| < 4.5\max(\sqrt{V_i}, \sqrt{V_j}) $. This stencil arrangement strategy can ensure stencil compactness and also enough particles are incorporated in each stencil when the particle distribution is not isotropic in the Lagrangian framework simulations. The stencil arrangement is summarized in Table \ref{tb:sten}. Unlike the conventional MLS-WENO-SPH method proposed by Avesani et al. \cite{avesani2014new}, where only the fourth-order reconstruction is devised, the present method provides up to the sixth-order reconstruction in a unified framework.

	\begin{table}[]
			\renewcommand{\arraystretch}{1.3}
	\caption{Stencil arrangements of central and directional stencils for different orders and their particle selection. } 	
		\begin{center}
		\begin{tabular}{@{}ccc@{}}
			\toprule
			Stencils                                       & Order                    & Particles \\ \midrule
			\multicolumn{1}{c|}{\multirow{3}{*}{Central}} & \multicolumn{1}{c|}{4th} &  $ j\in S_0 $, if $ \left| \boldsymbol{r}_{ij} \right| < 2.5\max(\sqrt{V_i}, \sqrt{V_j}) $       \\ \cmidrule(l){2-3} 
			\multicolumn{1}{c|}{}                         & \multicolumn{1}{c|}{5th} &  $ j\in S_0 $, if $ \left| \boldsymbol{r}_{ij} \right| < 3.2\max(\sqrt{V_i}, \sqrt{V_j}) $       \\ \cmidrule(l){2-3} 
			\multicolumn{1}{c|}{}                         & \multicolumn{1}{c|}{6th} &  $ j\in S_0 $, if $ \left| \boldsymbol{r}_{ij} \right| < 4\max(\sqrt{V_i}, \sqrt{V_j}) $      \\ \midrule
			Directional                                   & 3rd                      & $ j\in S_s $, if $ \pi (s-1)/4 \le \tan^{-1}(\frac{y_{ij}}{x_{ij}}) \le \pi s/4$ and $  \left| \boldsymbol{r}_{ij} \right| < 4.5\max(\sqrt{V_i}, \sqrt{V_j}) $ \\ \bottomrule
		\end{tabular}
		\end{center}
	\label{tb:sten}
	\end{table}
	
	\subsubsection {Smoothness indicators of candidate stencils}
	The smoothness indicator is usually employed to characterize the function smoothness, and this plays a significant role to ensure numerical stability and avoid artificial numerical oscillations. 
	The smoothness indicator \cite{jiang1996efficient} for the high-order approximation on the $ s $-th stencil is usually expressed as
	\begin{equation}
	\beta_s = \sum_{q=1}^n   \int_{V_i^{\prime}} \left( \mathcal{D}^q \tilde{u}_s(\xi,\eta)\right)^2 d\xi d\eta,
	\end{equation}
	where the integral domain  $ V_i^{\prime} $ is selected as a square with the edge length of $ 2 $ in the system of the non-dimensionalized variables $\xi$ and $\eta$. 
	Usually, the calculation of this equation is computationally expensive. Herein, we adopt a simplified formula of the smoothness indicator as
	\begin{equation}
	\begin{aligned}
	\beta_s = \sum_{q=1}^n   \int_{V_i^{\prime}} \left[  \mathcal{D}^q \left( \boldsymbol{N}^T \boldsymbol{b}^\prime \right)\right]^2  d\xi d\eta= \sum_{q=1}^n  \int_{V_i^{\prime}}   \sum_{i=1}^m \sum_{j=1}^m b^\prime_{(i)} b^\prime_{(j)} \left(  \mathcal{D}^q N_{(i)} \right) \left(  \mathcal{D}^q N_{(j)} \right)     d\xi d\eta \\
	=    \sum_{i=1}^m \sum_{j=1}^m b^\prime_{(i)} b^\prime_{(j)} \left( \sum_{q=1}^n   \int_{V_i^{\prime}}   \left(  \mathcal{D}^q N_{(i)} \right) \left(  \mathcal{D}^q N_{(j)} \right)     d\xi d\eta \right),
        \label{eq:sm_in}
	\end{aligned}
	\end{equation}
	where $b^\prime_{(i)}$ and $N_{(i)}$ denote the $i$-th element in $\boldsymbol{b}^\prime$ and $\boldsymbol{N}$, respectively. In Eq. (\ref{eq:sm_in}), the integral can be calculated by a 2D Gaussian quadrature, and the coefficients in the brackets can be calculated before simulation and stored beforehand. Unlike the approximate calculation of the smoothness indicator in \cite{avesani2014new}, we calculate the smoothness indicator strictly with its original definition.  
	
	\subsubsection {Nonlinear weighting strategy for the final reconstruction}
	\label{sect:teno_recon}
	The weighting strategy is to combine the field value approximations among a set of candidate stencils, and this is significant to ensure a robust and accurate  approximation of a field value. 
	A strong scale separation indicator is proposed in the work of \cite{fu2019very} to separate the non-smooth stencils from the smooth ones. The strong scale separation indicator $ \gamma_s $ of the $ s $th stencil is written as
	\begin{equation}
     \label{eq:gamma}
	\gamma_s = \dfrac{1}{(\beta_s +\epsilon)^6},
	\end{equation}
	where $ \epsilon=10^{-12} $ is used to evade the zero denominator. Given that $ \gamma_s $ functions of all candidate stencils are calculated,  the measured smoothness indicator $ \chi_0 $ for the large central stencil  is expressed as
	\begin{equation}
	\chi_0 = \dfrac{\gamma_0}{\sum_ {s=0} ^8 \gamma_s}. 
	\end{equation}
	Then, a cut-off parameter $ C_T $ is employed to determine whether to use the large central stencil. This parameter is related to the cut-off wavenumber of flow scales, and the selection of the value is discussed in \cite{fu2019targeted}\cite{fu2016family}. At last, the cut-off function  $ \delta_0 $ is determined by comparing to $ C_T $, written as
	\begin{equation}
    \label{eq:sig0}
	\delta_0= \begin{cases}0, & \text { if } \chi_0<C_T, \\
	1, & \text { otherwise }.\end{cases}
	\end{equation}
	The value of $ C_T $ is usually from $ 10^{-7} $ to $ 10^{-5} $ \cite{fu2016family}. In this work, for the fourth-order (O4) reconstruction on the large central stencil, $ C_T=10^{-5} $; for fifth-order (O5) reconstruction on the large central stencil, $ C_T=10^{-6} $; for sixth-order (O6) reconstruction, $ C_T=10^{-7} $. 
	The cut-off function  $ \delta_0 $ determines whether to use the large central stencil for the final reconstruction. The detailed stencil selection strategy is as follows: 
	
	(1) If the large central stencil is activated, i.e., $ \delta_0=1 $, then all the other small directional stencils will not be used, then the reconstructed variable $ \hat{\mathbf{U}}_i(\xi, \eta) $ is equal to the MLS approximation of the $ 0 $th stencil, $\hat{\mathbf{U}}_i^{(0)}(\xi, \eta) $, written as
	\begin{equation}
	\hat{\mathbf{U}}_i(\xi, \eta) = \hat{\mathbf{U}}_i^{(0)}(\xi, \eta).
        \label{eq:comb_s1}
	\end{equation}
	
	(2) If the large scale is not activated, i.e., $ \delta_0=0 $, then the other small directional stencils will be used. The reconstructed variable $ \hat{\mathbf{U}}_i(\xi, \eta) $ will be the average of the activated small stencils, expressed as
	\begin{equation}
	\hat{\mathbf{U}}_i(\xi, \eta) =\sum_{s=1}^8 \omega_s \hat{\mathbf{U}}_i^{(s)}(\xi, \eta),
        \label{eq:comb_s2}
	\end{equation}
	where $ \omega_s $ is the corresponding weight for each candidate stencil, written as
	\begin{equation}
	\omega_s = \dfrac{\gamma_s}{\sum_ {s=1} ^8 \gamma_s}. 
	\end{equation}

    This criterion ensures that the high-order reconstruction  can be achieved locally when the MLS data construction in the central stencil is smooth, and also evade artificial numerical oscillations when crossing discontinuities. 

	\subsection{High-order discretization for gradient operator in smooth regions}
	In compressible flows, the particle distribution is usually not isotropic, and it leads to the inconsistency of the particle discretization, i.e., the accuracy order of the gradient approximation, e.g., $ \sum_j^N  2 \tilde{\mathbf{G}}_{i j} \cdot \nabla W_{i j}V_i V_j $ in Eq. (\ref{eq:ale_vila1}), is very low. With the non-isotropic particle distributions, even $ \tilde{\mathbf{G}}_{i j} $ is exact, the gradient approximation is  inaccurate under this circumstance. 
	
	Indeed, the components of the gradient operator $ \frac{\hat \partial }{\partial x} $ and  $ \frac{\hat \partial}{\partial y} $ have already been calculated in the MLS step. Note that 
	$  \frac{\partial \mathbf{f}_i}{\partial x} = \frac{\partial \mathbf{f}_i}{\mathbf{U}_i} \frac{\partial \mathbf{U}_i}{\partial x} $ and $  \frac{\partial \mathbf{g}_i}{\partial y} = \frac{\partial \mathbf{g}_i}{\mathbf{U}_i}  \frac{\partial \mathbf{U}_i}{\partial y} $, and the reconstructed field $ \mathbf{\hat U}_i $ can be derived with the merit of MLS method. Then
	\begin{equation}
	\frac{\hat \partial}{\partial x}  \mathbf{f}_i = \mathbf{A}_i \frac{\partial  {\mathbf{\hat U}_i} }{\partial x}, \  \frac{\hat \partial }{\partial y} \mathbf{g}_i = \mathbf{B}_i \frac{\partial  {\mathbf{\hat U}_i} }{\partial y},
	\label{eq:sm_gd}
	\end{equation}

	and the ALE term in Eq. (\ref{ale}) is written as
	\begin{equation}
	\nabla \cdot (  \mathbf{U}_i \otimes \boldsymbol{w}_i ) = \boldsymbol{w}_i\cdot \nabla \mathbf{U}_i = \boldsymbol{w}_i\cdot (\frac{\partial  {\mathbf{\hat U}_i} }{\partial x}, \frac{\partial  {\mathbf{\hat U}_i} }{\partial y}),
	\label{eq:sm_gd1}
	\end{equation}	
	where $ \frac{\partial  {\mathbf{\hat U}_i} }{\partial x} $ and $ \frac{\partial  {\mathbf{\hat U}_i} }{\partial y} $ are easy to obtain when   $ \mathbf{\hat U}_i $ is constructed with the MLS method in Eq. (\ref{eq:msl1}). They can be written as
    \begin{equation}
    \frac{\partial  {\mathbf{\hat U}_i} }{\partial x}=\left(b^{\rho}_{(1)}, b^{\rho u}_{(1)}, b^{\rho v}_{(1)}, b^{E}_{(1)}\right)^T, \  \frac{\partial  {\mathbf{\hat U}_i} }{\partial y}=\left(b^{\rho}_{(2)}, b^{\rho u}_{(2)}, b^{\rho v}_{(2)}, b^{E}_{(2)}\right)^T,
    \end{equation}    
    where $b^{\psi}_{(i)}$ denotes the $i$-th element of vector $\boldsymbol{b}$ in Eq. (\ref{eq:msl1}) for the conservative variable $\psi$ in $\mathbf{\hat U}_i$. The Jacobian matrix $ \mathbf{A}_i $ and $ \mathbf{B}_i $ can also easily be obtained using local field values of particle $ i $. Eq. (\ref{eq:sm_gd}) can achieve genuine high-order accuracy when the local flow field is smooth. This kind of construction for the gradient operator is usually non-conservative. This issue may not be that critical, since the numerical discretization errors as well as the conservation errors will vanish rapidly with the high-order convergence. Indeed, to date, no gradient construction method for SPH can achieve both high-order accuracy and exact conservation at the same time.

\subsection{Unified high-order framework for both smooth and non-smooth flows}

	In the  section above, the high-order discretization of gradient operator is proposed, which is only applicable in  smooth regions. When the solution field is not smooth, i.e., with shock discontinuities or high-wavenumber under-resolved flow scales, Eq. (\ref{eq:sm_gd})  will lead to artificial numerical oscillations. As such, in these regions, the Vila's SPH framework should be employed for numerical stability based on an appropriate Riemann solver, i.e., Eq.~(\ref{eq:Riemann_flux}). In this work, a unified framework will be proposed, which restores the high-order  MLS reconstruction in smooth regions and the nonlinear shock-capturing TENO scheme in the non-smooth regions.
	
	Herein, the key question is how to determine the smooth and non-smooth regions. In the section above, we employ a strong scale separation indicator $ \gamma_0 $, Eq. (\ref{eq:gamma}), and a cut-off function $ \delta_0 $, Eq. (\ref{eq:sig0}), to determine whether the large central stencil should be activated or not. Based on the functions of $ \delta_0 $, the smooth region indicator $ \phi $ is further defined to indicate if the corresponding particle is in the smooth region or not, and $ \phi=1 $ signifies the smooth region and $ \phi=0 $ the non-smooth region. Specifically, $ \phi_i $ is determined using $ \delta^j_0 $ of its neighboring particles $ j $, written as
	\begin{equation}
	\phi_i= \begin{cases}0, & \exists\text { particle } j \text { within radius } 8\Delta x\text {, } \delta^j_0=0, \\
	1, & \text { otherwise,}\end{cases}
	\end{equation}
	where the first condition means that if there exists a particle $ j $ within the radius $ 8\Delta x $ of particle $ i $ featuring $ \delta^j_0=0 $, then particle $ i $ is defined to be in the non-smooth region, otherwise it is in the smooth region. Correspondingly, the final discretized form of the governing equations can be given, as
	\begin{equation}
    \label{eq:uni}
	\dfrac{\tilde{d} V_i \mathbf{U}_i}{d t}= \begin{cases}-\sum_j^N  2 (\tilde{\mathbf{G}}_{i j}-\mathbf{H}_i) \cdot \mathbf{M}_{i}^{-1}\nabla W_{i j}V_i V_j, & \text { if } \phi_i=0, \\
	\left[ -\left(\mathbf{A}_i  \frac{\partial  {\mathbf{\hat U}_i} }{\partial x} + \mathbf{B}_i\frac{\partial  {\mathbf{\hat U}_i} }{\partial y} \right)+\boldsymbol{w}_i\cdot (\frac{\partial  {\mathbf{\hat U}_i} }{\partial x}, \frac{\partial  {\mathbf{\hat U}_i} }{\partial y}) \right]   V_i, & \text { otherwise.}\end{cases}
	\end{equation}
    This separation of smooth and non-smooth flow fields can, on the one hand, restore the genuine high-order discretization accuracy in smooth regions, and on the other hand, be highly computationally efficient. In Eq. (\ref{eq:uni}), for $\phi_i=0$, the looping calculation of neighboring particles is usually time-consuming. While for $\phi_i=1$, only simple matrix calculations are needed given the obtained coefficients of MLS construction in the central stencil, and thus the tremendous looping calculation of neighboring particles is not required. 
    
    The above unified numerical method will be named as MLS-TENO-SPH hereafter. For the time integration, the second-order TVD Runge-Kutta method \cite{gottlieb1998total} is used, and the time step is set as $\Delta t = \min(0.2{h_i}/{c_i})$.

	\section{Numerical validations}
	In this section, several benchmark cases are used to validate the performance of the present MLS-TENO-SPH framework, i.e., the high-order accuracy and the shock-capturing capabilities. In most cases, the results from the MLS-WENO-SPH method \cite{avesani2014new}, which is designed to feature the fourth-order reconstruction accuracy, will be provided for comparisons.
 
	\subsection{Accuracy order test}
	In this section, the performance of the TENO and SPH reconstructions for the field value and  the gradient operator is investigated. Then, the linear advection equation is employed to verify the high-order convergence of the present framework.

	\subsubsection{Reconstruction order verifications}
	The TENO reconstruction method is used to reconstruct the smooth field function
	\begin{equation}
	f(x,y)= e^{-0.2(x+y)} \left( \cos\left( 4x\right) +\sin\left( 4y\right)  \right), 
	\end{equation}
	where $ (x,y) $ denotes the particle coordinate. A uniform particle distribution is set up first, and the staggered reconstruction particles are deployed with the same resolution. Four different resolutions, $ \Delta x=1/75 $,  $ \Delta x=1/150 $,  $ \Delta x=1/300 $ and  $ \Delta x=1/600 $, are used to test the TENO reconstruction order. Also, two different kinds of particle distribution, structured (lattice) and unstructured (disorder), are considered. In detail, for the structured particle distribution, a lattice distribution with an initial spacing of $\Delta x$ is used; for the unstructured particle distribution, the particles are disordered randomly away from a lattice particle distribution with the variance of $0.5\Delta x$. 
	The measured numerical discretization error with $ N $ particles is as 
	\begin{equation}
	L_2=\sqrt{\sum_{j=1}^N\left|f_r\left(x_j, y_j\right)-f_e\left(x_j, y_j\right)\right|^2/N },
	\end{equation}
	where $ f_r  $ and $ f_e $ denote the reconstructed and exact values, respectively. 

\begin{figure}[htbp]
		\centering
		\includegraphics[width=0.7\textwidth]{ 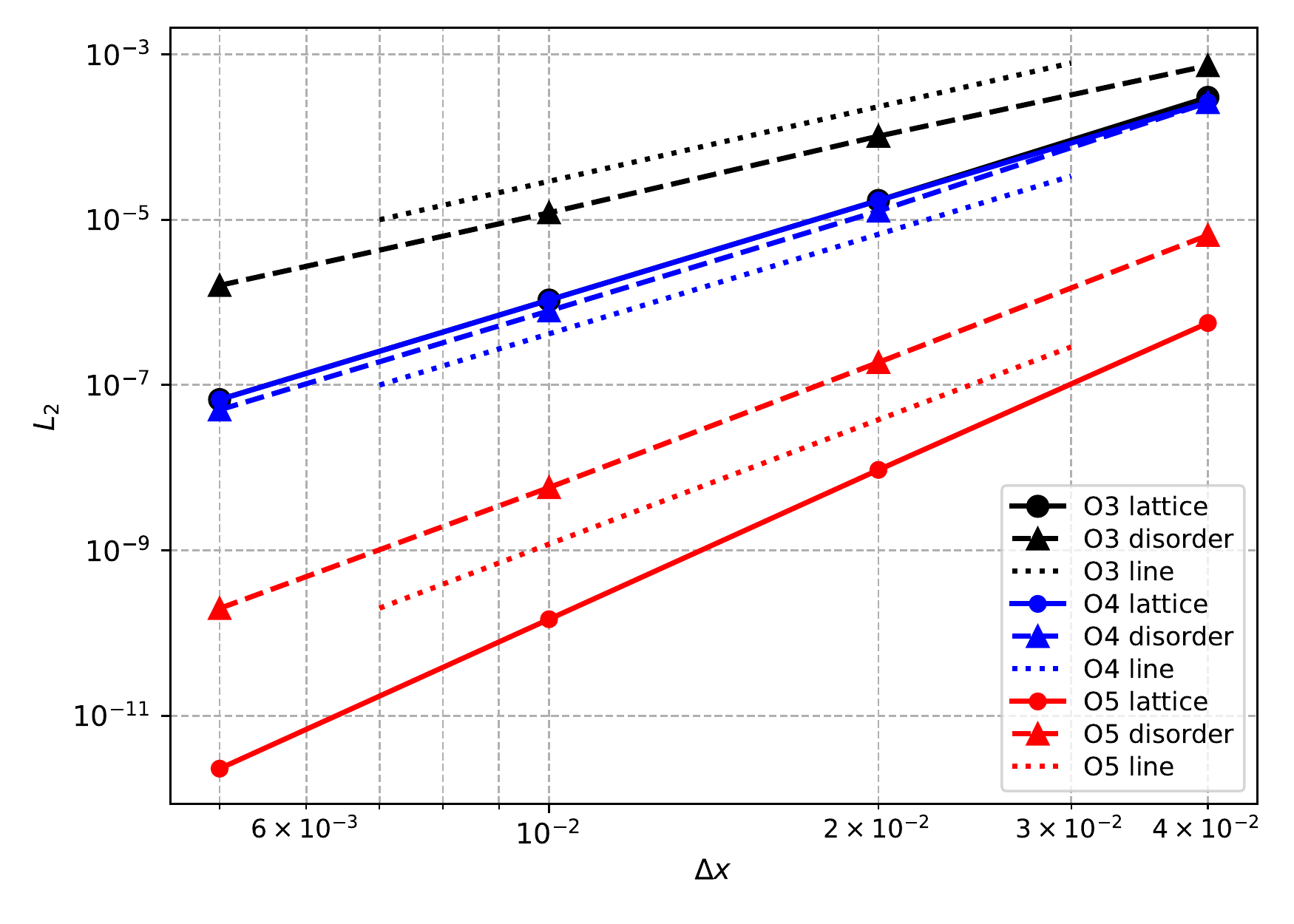}
		\caption{The convergence statistics for reconstructing the functional $f(x,y)$ with the third- (O3) fourth- (O4) and fifth- (O5) order MLS-TENO reconstructions, where both the structured (lattice) and unstructured (disorder) particle distributions are considered. }
		\label{ord_n}
	\end{figure}

\begin{figure}[htbp]
		\centering
		\includegraphics[width=0.7\textwidth]{ 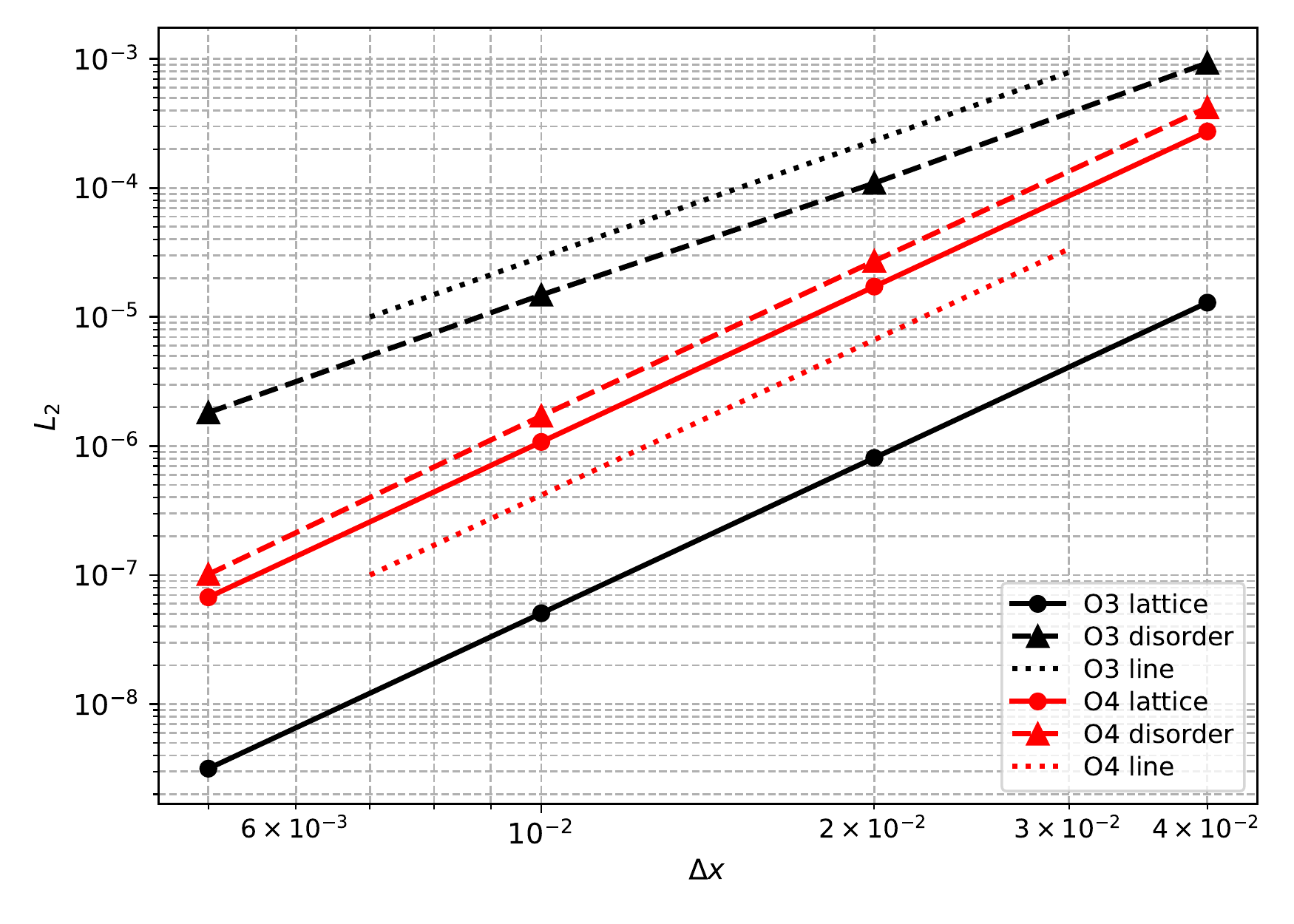}
		\caption{The convergence statistics for the divergence approximation of $  \nabla \cdot \left(  f(x,y),f(x,y) \right)  $ with the  fourth- and fifth-order  MLS-TENO reconstructions, for which the expected divergence approximation order are third- (O3) and fourth- (O4) order, respectively. Both the structured (lattice) and unstructured (disorder) particle distributions are considered. }
		\label{ord_g_n}
	\end{figure}

    Fig. \ref{ord_n} displays the convergence statistics with the TENO reconstruction based on MLS method (MLS-TENO) for the given smooth function with respect to the third- (O3), fourth- (O4) and fifth- (O5) order. The expected accuracy order is obtained for all the considered TENO schemes. It is further shown that the lattice particle distribution usually leads to smaller discretization errors.
	Using the same field function, the divergence approximation order of $  \nabla \cdot \left(  f(x,y),f(x,y) \right)  $ is tested. The expected divergence approximation order is one order lower than the MLS data reconstruction order due to the derivative operation, i.e., the expected divergence approximation order of MLS-TENO (O4) and MLS-TENO (O5) are third- (O3) and fourth- (O4) order, respectively. 
 As shown in Fig. \ref{ord_g_n}, for both particle distributions, the expected accuracy order is achieved with the present MLS-TENO methods.


	To validate the shock-capturing properties of the proposed scheme, the non-smooth function is considered as follows
	\begin{equation}
	f(\left| \boldsymbol{r} \right| )= \begin{cases} 2.5, \qquad \text { if}\  \left| \boldsymbol{r} \right|<0.2, \\
	2, \qquad \text { if}\  0.2 \le \left|\boldsymbol{r} \right|<0.4, \\
	\left| \boldsymbol{r} \right| + 1.6, \qquad \text { if}\  0.4 \le \left|\boldsymbol{r} \right|<0.6, \\
	-(\left| \boldsymbol{r} \right|-0.6)^2+2.2, \qquad \text { otherwise}. \end{cases}
	\end{equation}

\begin{figure*}[htbp]
		\centering
		
		\subfigure[Latticed particle distribution]{
			\includegraphics[width=0.46\linewidth]{ 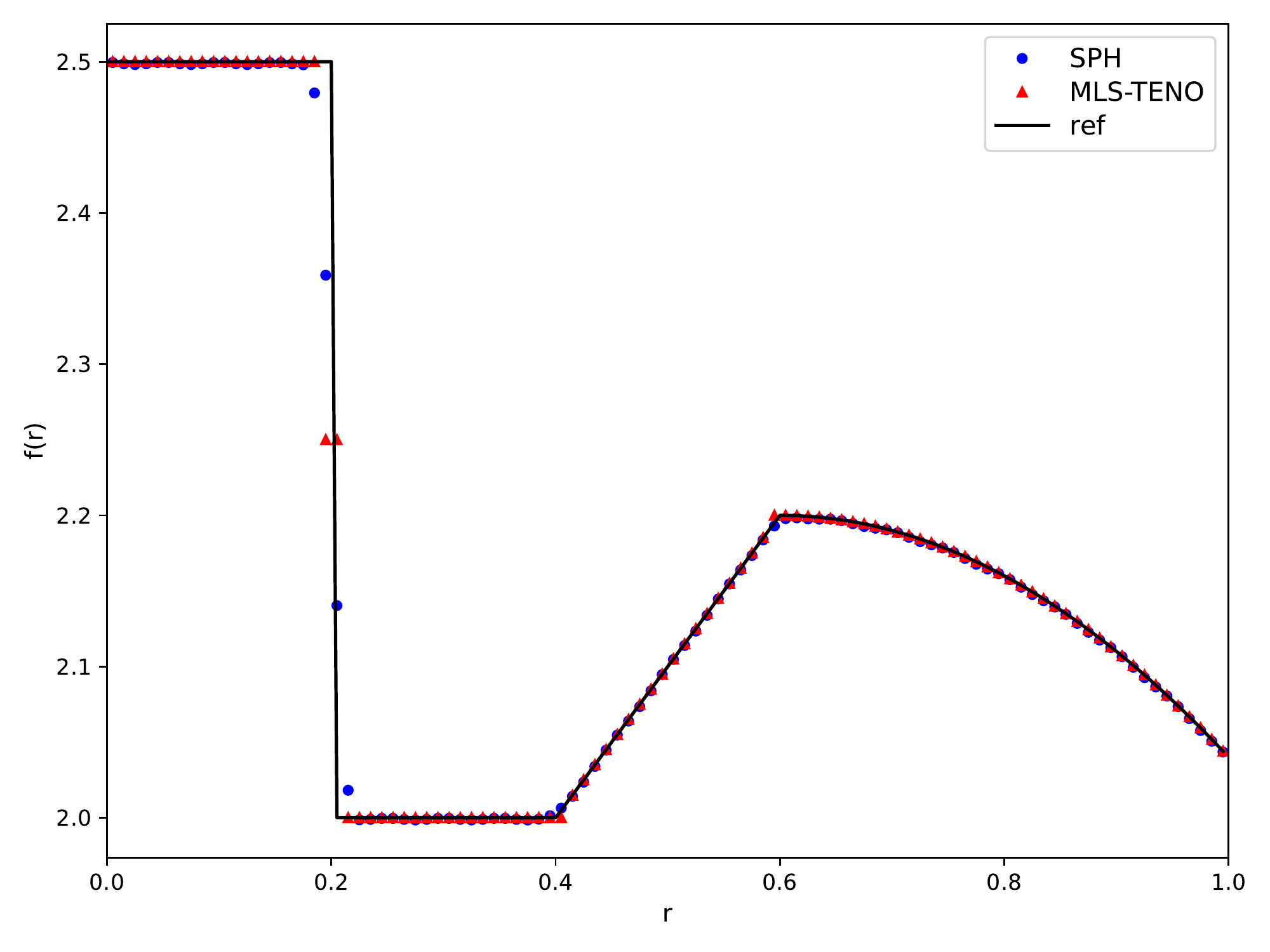}}
		\subfigure[Disordered particle distribution]{
			\includegraphics[width=0.46\linewidth]{ 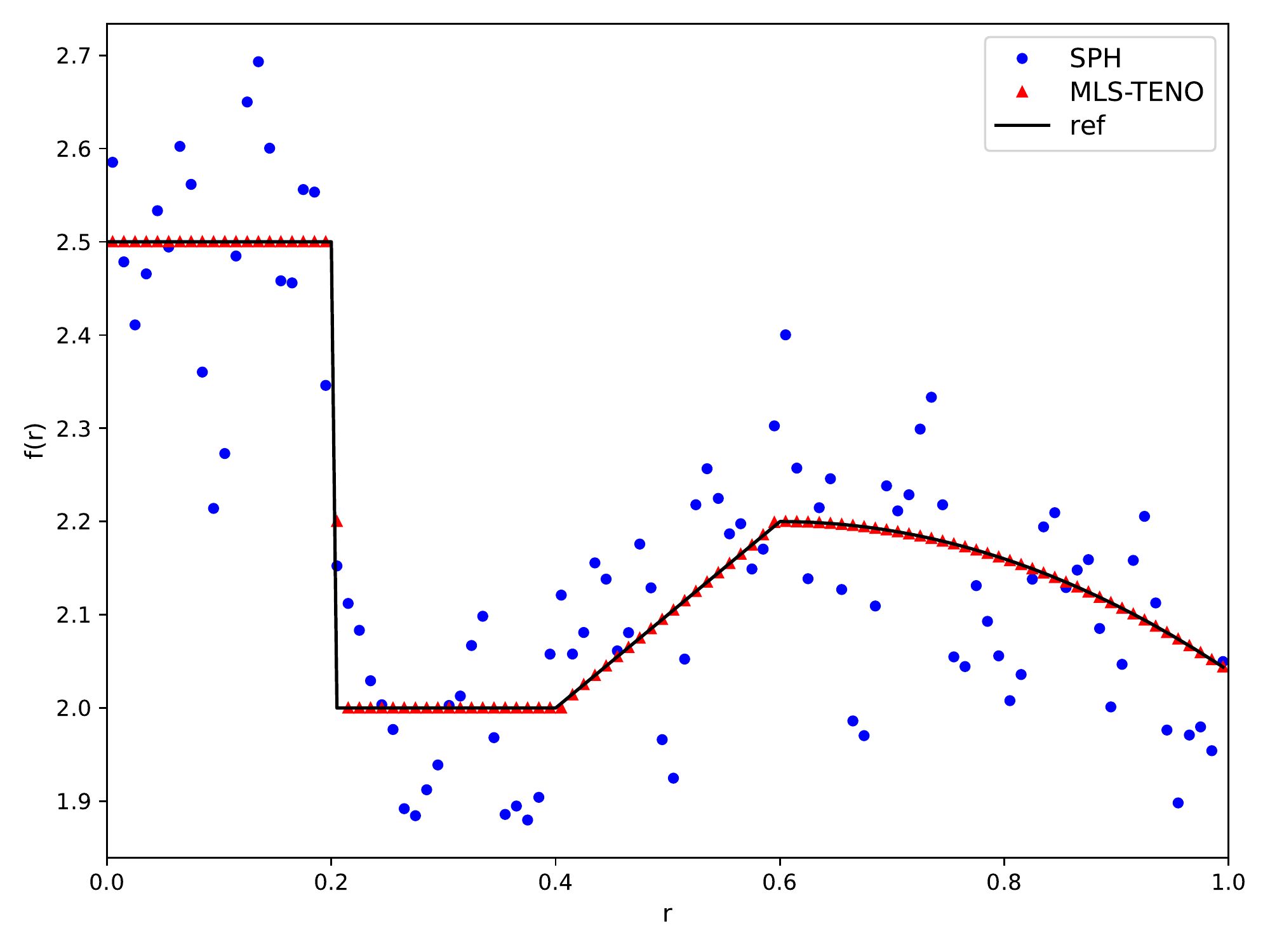}}
		
		\caption{Reconstructed field values using MLS-TENO and SPH kernel for latticed and disordered particle distributions when the strong discontinuity presents.}
		\label{rec_f}
	\end{figure*}

	The comparison between MLS-TENO  and the conventional SPH reconstruction is shown in Fig. \ref{rec_f}. In terms of the latticed particle distribution, the performance of MLS-TENO is slightly better than SPH. However, when the initial particle distribution is in disorder, the SPH reconstructed solution deviates significantly from the reference, which is due to the kernel inconsistency of SPH for this case. On the other hand, the MLS-TENO scheme performs quite well even with the discontinuity presented in the solution.

	\subsubsection{Linear advection evolution}
	A 2D linear advection problem is used to validate the convergence order of the present scheme. The 2D linear advection equation is written as
	\begin{equation}
	\dfrac{\partial u}{\partial t} + \dfrac{\partial u}{\partial x} + \dfrac{\partial u}{\partial y}=0, 
	\end{equation}
	and the initial condition is
	\begin{equation}
	u(x,y,0)= \sin(2\pi x ) \sin(2\pi y).
	\end{equation}
	The periodic boundary condition is implemented in this case and the simulation time is set as $ t=0.5 $. The timestep is set as small enough to eliminate the time integration error.
	
\begin{figure}[htbp]
		\centering
		\includegraphics[width=0.7\textwidth]{ 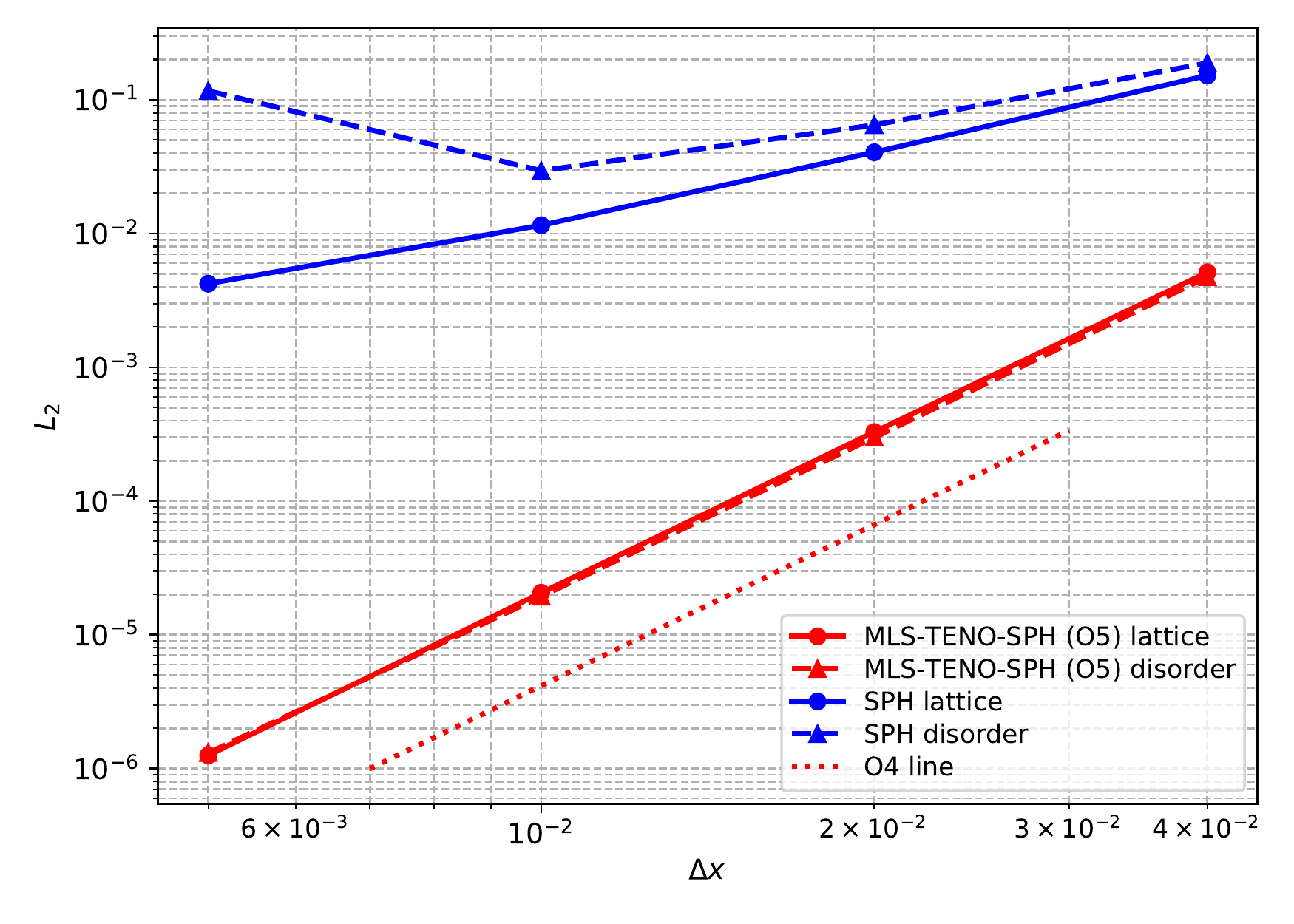}
		\caption{Convergence statistics of the 2D linear advection problem with smooth field values, where the MLS-TENO-SPH (O5) framework is employed for the simulations. The statistics simulated using the conventional Vila's SPH method with fourth-order reconstruction for field variables are also shown for comparisons.} 
		\label{ord_adv}
	\end{figure}

	The convergence statistics for resolutions $ \Delta x=1/25 $,  $ \Delta x=1/50 $,  $ \Delta x=1/100 $ and  $ \Delta x=1/200 $ are plotted in Table \ref{tb:converg} and Fig. \ref{ord_adv}. In this test, the MLS-TENO-SPH (O4 and O5) framework is employed for the simulations.  Two different kinds of particle distribution, i.e., latticed and disordered, are used. Specifically, for the disordered particle distribution, we disorder particles randomly away from a lattice particle distribution with the variance of $0.3\Delta x$, referring to  \cite{nasar2021high}. The conventional Vila's SPH method with fourth-order reconstruction for field variables are also shown for comparisons. 

    \textcolor{black}{The expected accuracy orders of the MLS-TENO-SPH (O4 and O5) framework with respect to the first-order derivative are third- (O3) and fourth- (O4) order, respectively, in the smooth regions. In this table, for the lattice particle distribution, MLS-TENO-SPH (O4) even has a higher convergence  rate, fourth order, than the expected third order, and MLS-TENO-SPH (O4) almost has the same convergence  rate with MLS-TENO-SPH (O5). For the disordered particle distribution, the convergence  rate is slightly lower than that for lattice particle distribution, where the reduced convergence  rate may be attributed to the non-conservation property of the high-order MLS data reconstruction,  but their accuracy orders are all higher than that of the conventional Vila's SPH method. }
	In comparison, the conventional Vila's SPH framework cannot achieve the expected fourth-order convergence rate even for lattice particle distribution, owing to the low accuracy of the discretized gradient operator approximation. For the disordered particle distribution, the error even becomes larger when $ \Delta x$ is reduced  in the Vila's SPH framework.

\begin{table}[]
			\renewcommand{\arraystretch}{1.3}
	\caption{\textcolor{black}{Convergence tests for the case of linear advection equation, with Vila's SPH framework and present MLS-TENO-SPH framework. Herein, the lattice particle distribution and disordered particle distribution (generated by moving each particle randomly away from a lattice particle distribution with the variance of $0.3\Delta x$) are used for the tests.} } 	
 \begin{center}
\begin{tabular}{p{2cm} p{1cm} p{2cm}p{2cm}p{1.5cm}p{2.5cm}p{1.5cm}}
\toprule [1.2 pt]
                  & $\Delta x$ & \begin{tabular}[c]{@{}l@{}}Particle\\Number\end{tabular} & $L_2$, lattice & $\mathcal{O}(L_2)$, disorder & $L_2$, disorder & $\mathcal{O}(L_2)$, disorder \\ \hline
\multirow{4}{*}{SPH} &  0.04  &  625   & 1.52E-1 &   --  & 1.88E-1 &   --       \\
                  &   0.02   &  2500   &  4.05E-2  & 1.91   & 6.47E-2 &   1.54        \\
                  &    0.01  &  10000   & 1.15E-2 &  1.81   & 2.95E-2 &   1.13       \\
                  &    0.005 &  40000   & 4.21E-3 &  1.45   & 1.16E-1 &   -1.54       \\\hline
\multirow{4}{*}{\begin{tabular}[c]{@{}l@{}}MLS-TENO\\-SPH (O4)\end{tabular}} 
                    &   0.04   &  625  &  1.77E-3   &   -- & 8.95E-3 &   --     \\
                  &   0.02   &  2500  &    1.12E-4   &  3.98   & 5.76E-4 &   3.95      \\
                  &   0.01   &  10000   &  6.99E-6   &  4.00 & 6.82E-5 &   3.07      \\
                  &    0.005  &  40000  &  4.04E-7  &  4.11 & 1.54E-5 &   2.14     \\ \hline
\multirow{4}{*}{\begin{tabular}[c]{@{}l@{}}MLS-TENO\\-SPH (O5)\end{tabular}} 
                    &   0.04   &  625  &  5.12E-3   &   --  & 4.78E-3 &   --    \\
                  &   0.02   &  2500  &    3.28E-4   &  3.96 & 2.99E-4 &   3.98     \\
                  &   0.01   &  10000   &  2.06E-5   &  3.99 & 1.93E-5 &   3.94      \\
                  &    0.005  &  40000  &  1.25E-6  &  4.03 & 1.31E-6 &   3.88      \\ 
\bottomrule [1.2 pt]
\end{tabular}
\label{tb:converg}
\end{center}
\end{table}

	\subsection{Circular blast wave problem}
	The conventional 2D circular blast wave problem is usually used to investigate the performance of shock-capturing schemes in the vicinity of discontinuities. The initial condition is given as
	\begin{equation}
	(\rho, u, v, p)(\boldsymbol{r}, 0)= \begin{cases}\left(1, 0,0, 1 \right), & \text { if }\left|\boldsymbol{r} \right| \leqslant 0.5, \\ \left( 0.125, 0,0, 0.1 \right), & \text { otherwise, }\end{cases}
	\end{equation}
	where $ |\boldsymbol{r}| $ denotes the radius from the center. In this case, a circular particle distribution is deployed with the particle spacing $ \Delta x=0.01 $, and the whole computational domain is a square with the edge  length equal to $ 3 $. The simulation time is set as $ 0.2 $. Considering that the MLS-WENO-SPH method is fourth-order, a fourth-order MLS approximation (O4) is used for the large central stencil for a fair comparison. 
	
\begin{figure*}[htbp]
		\centering
		
		\subfigure[Density (Eulerian)]{
			\includegraphics[width=0.45\linewidth]{ 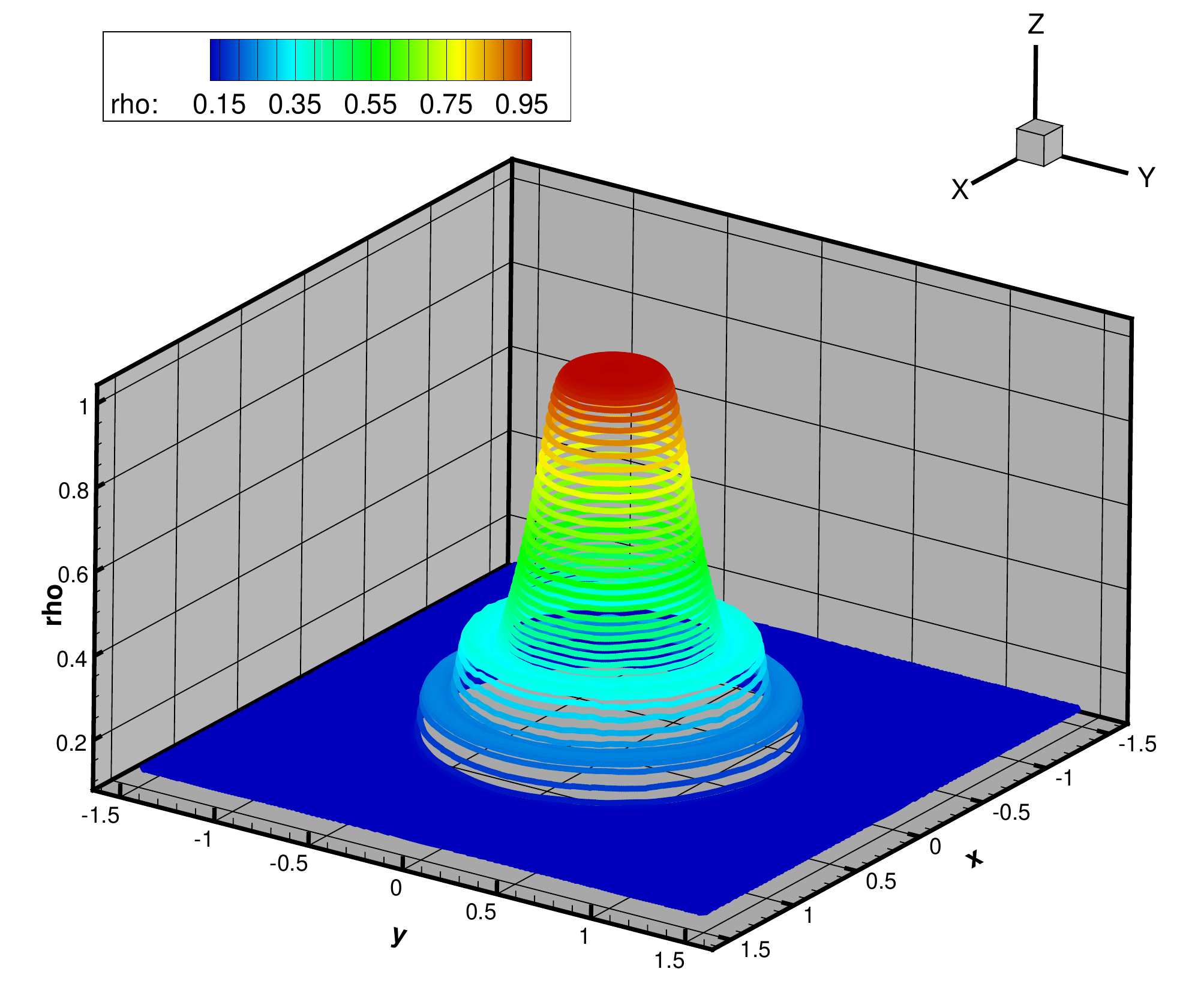}}
		\subfigure[Density (Lagragian)]{
			\includegraphics[width=0.45\linewidth]{ 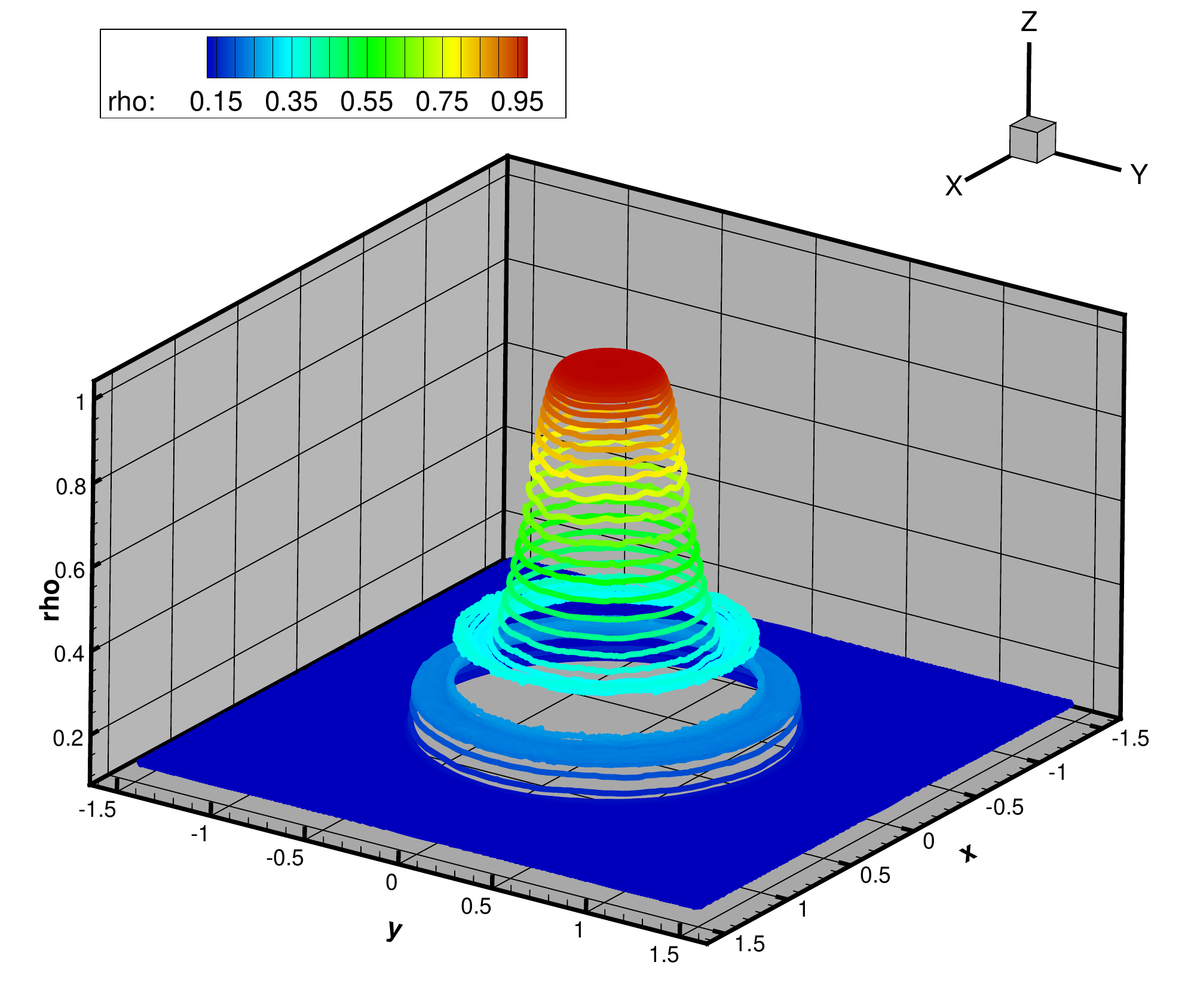}}
		\subfigure[Pressure (Eulerian)]{
			\includegraphics[width=0.45\linewidth]{ 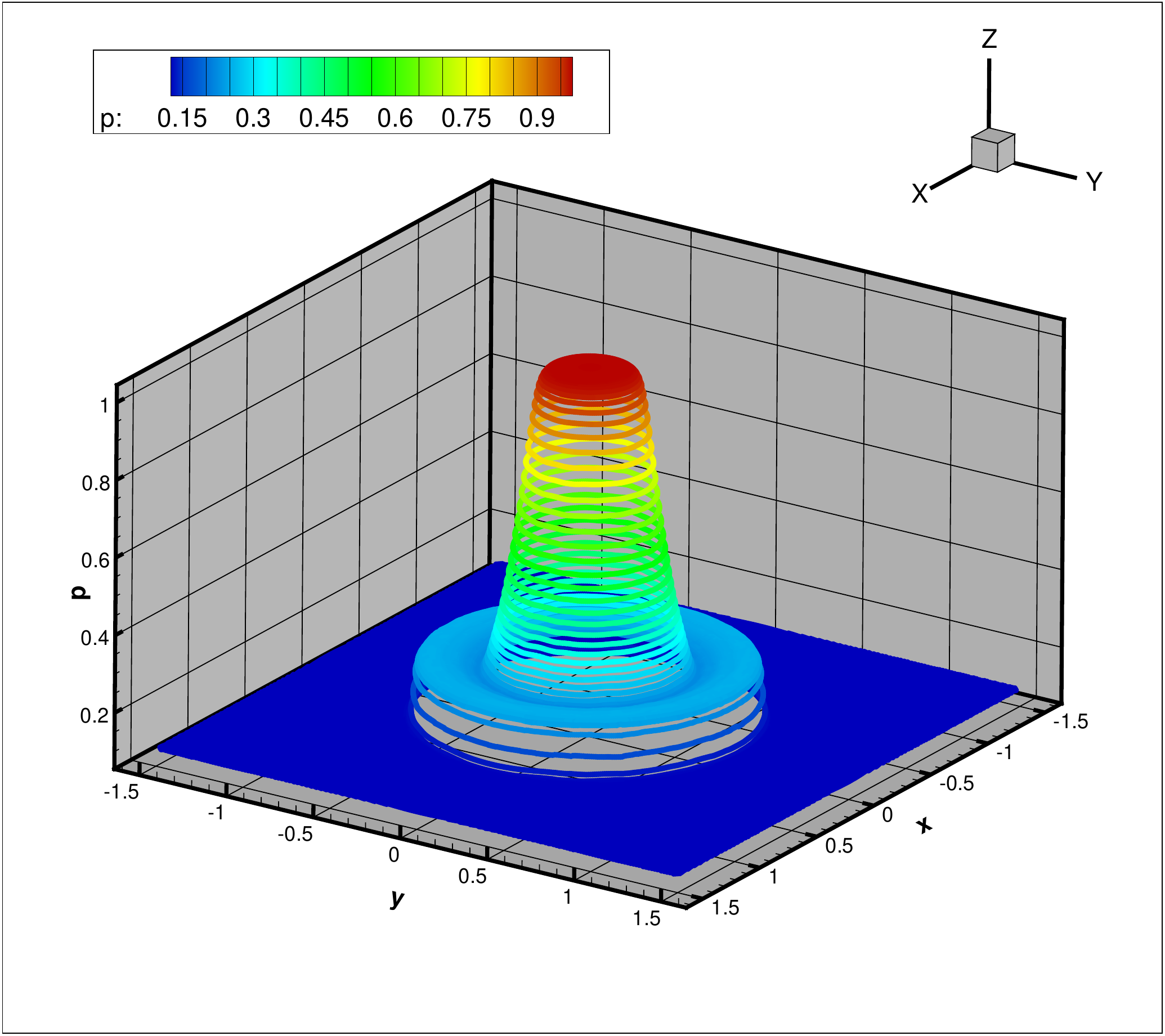}}
		\subfigure[Pressure (Lagragian)]{
			\includegraphics[width=0.45\linewidth]{ 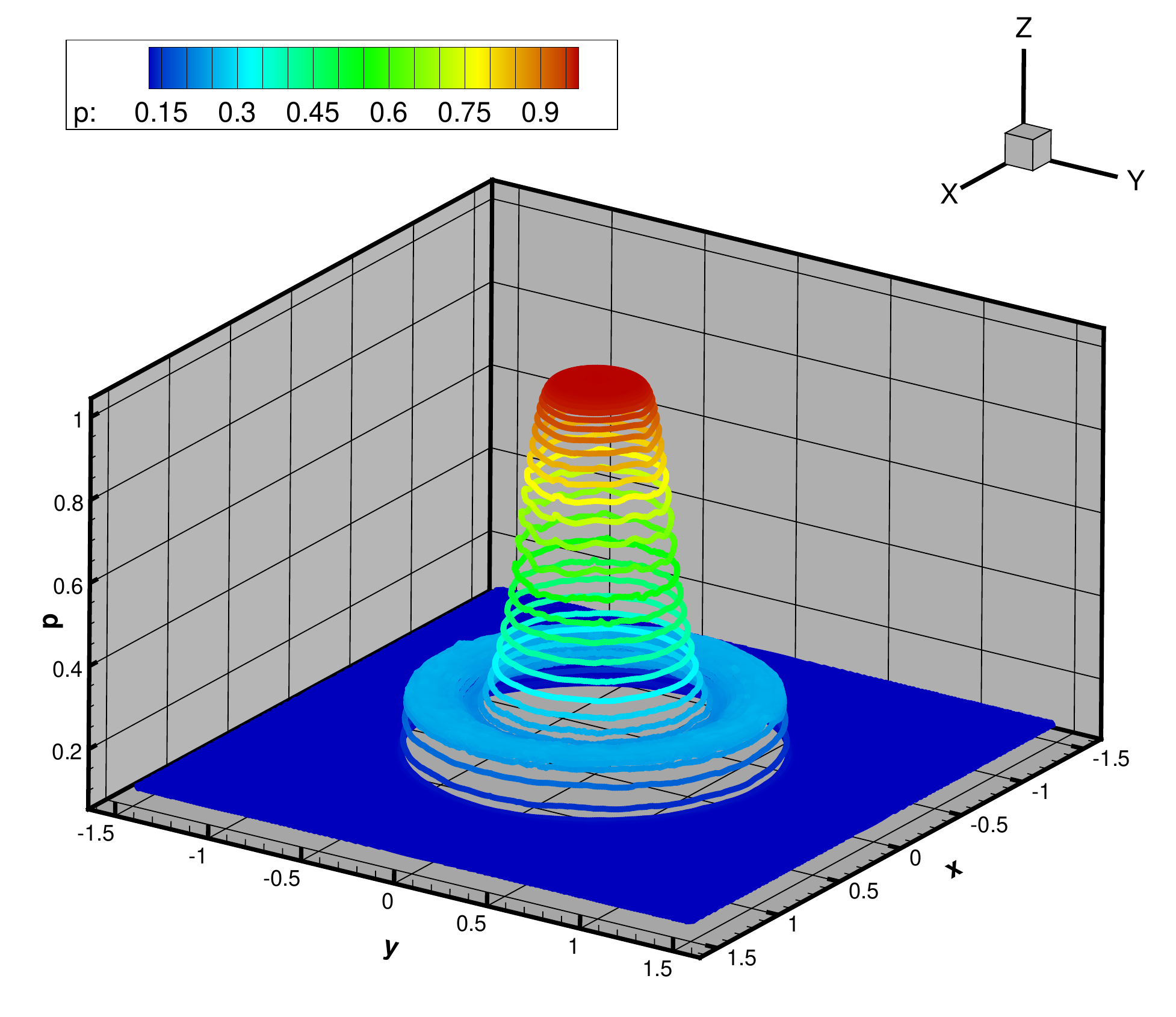}}
		
		\caption{Circular blast wave problem: density and pressure fields simulated with the present MLS-TENO-SPH(O4) method in the Eulerian and Lagrangian frameworks. }
		\label{cnt_bw}
	\end{figure*}

\begin{figure*}[htbp]
		\centering
		
		\subfigure[Density (Eulerian)]{
			\includegraphics[width=0.45\linewidth]{ 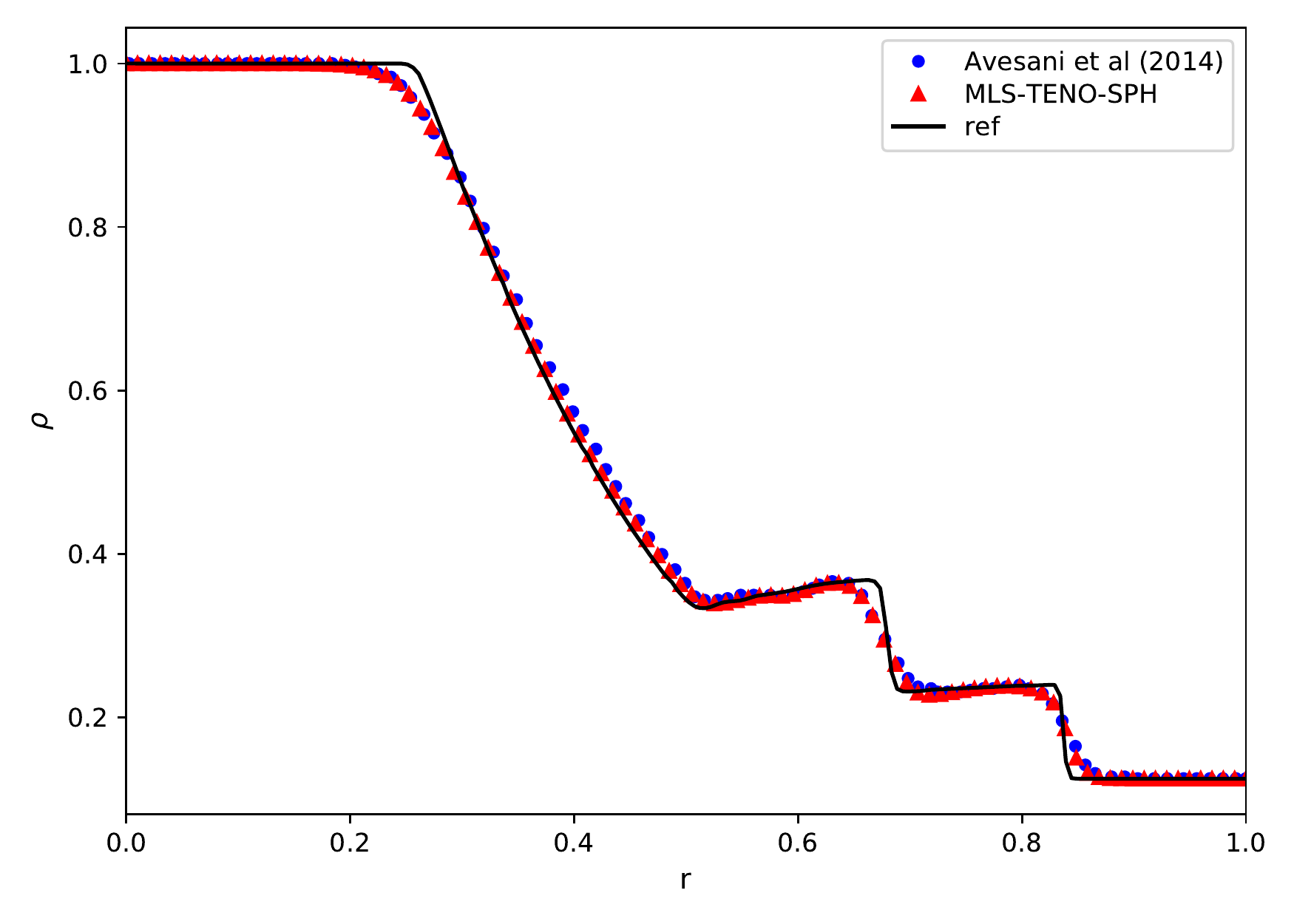}}
		\subfigure[Density (Lagragian)]{
			\includegraphics[width=0.45\linewidth]{ 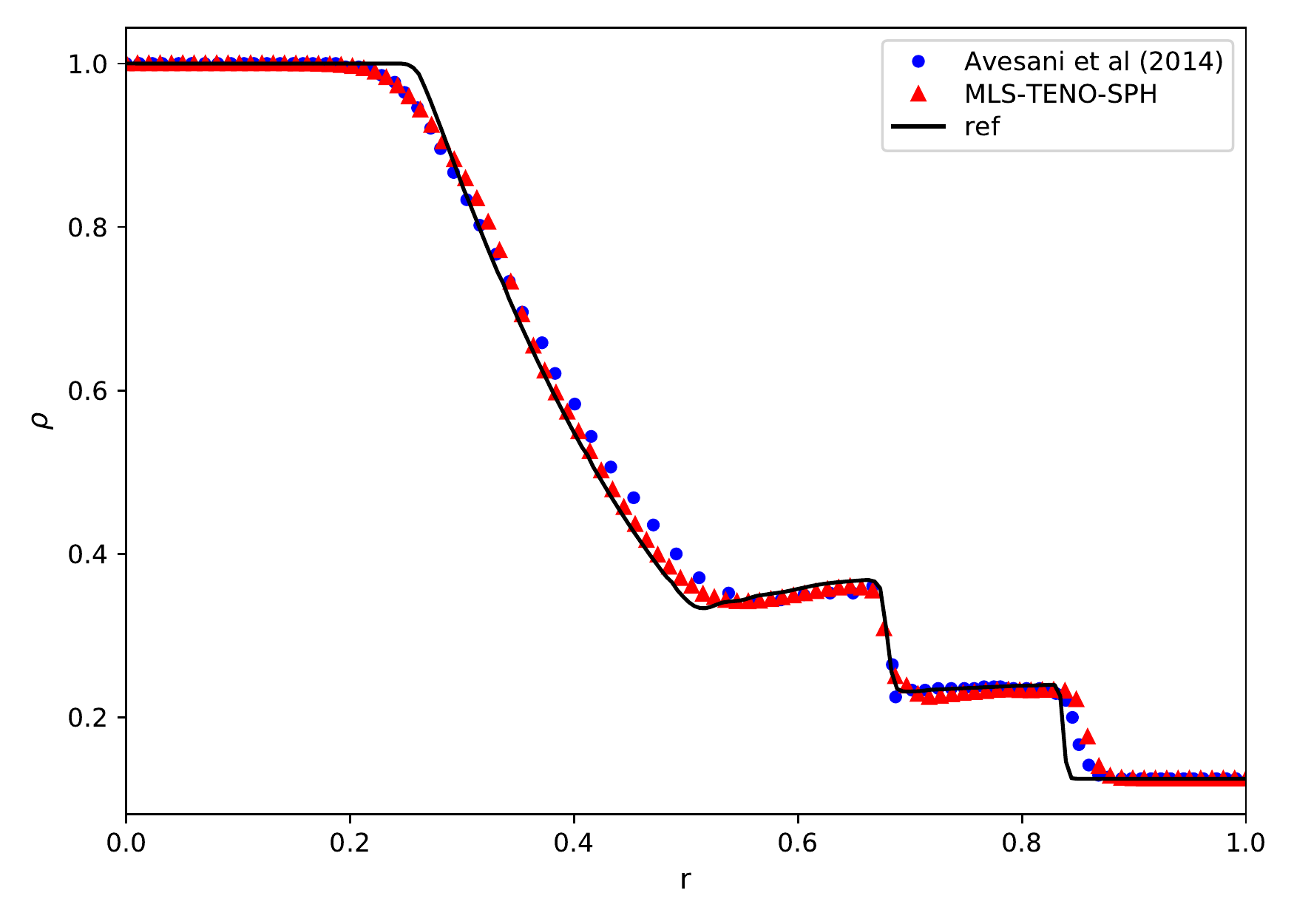}}
		\subfigure[Pressure (Eulerian)]{
			\includegraphics[width=0.45\linewidth]{ 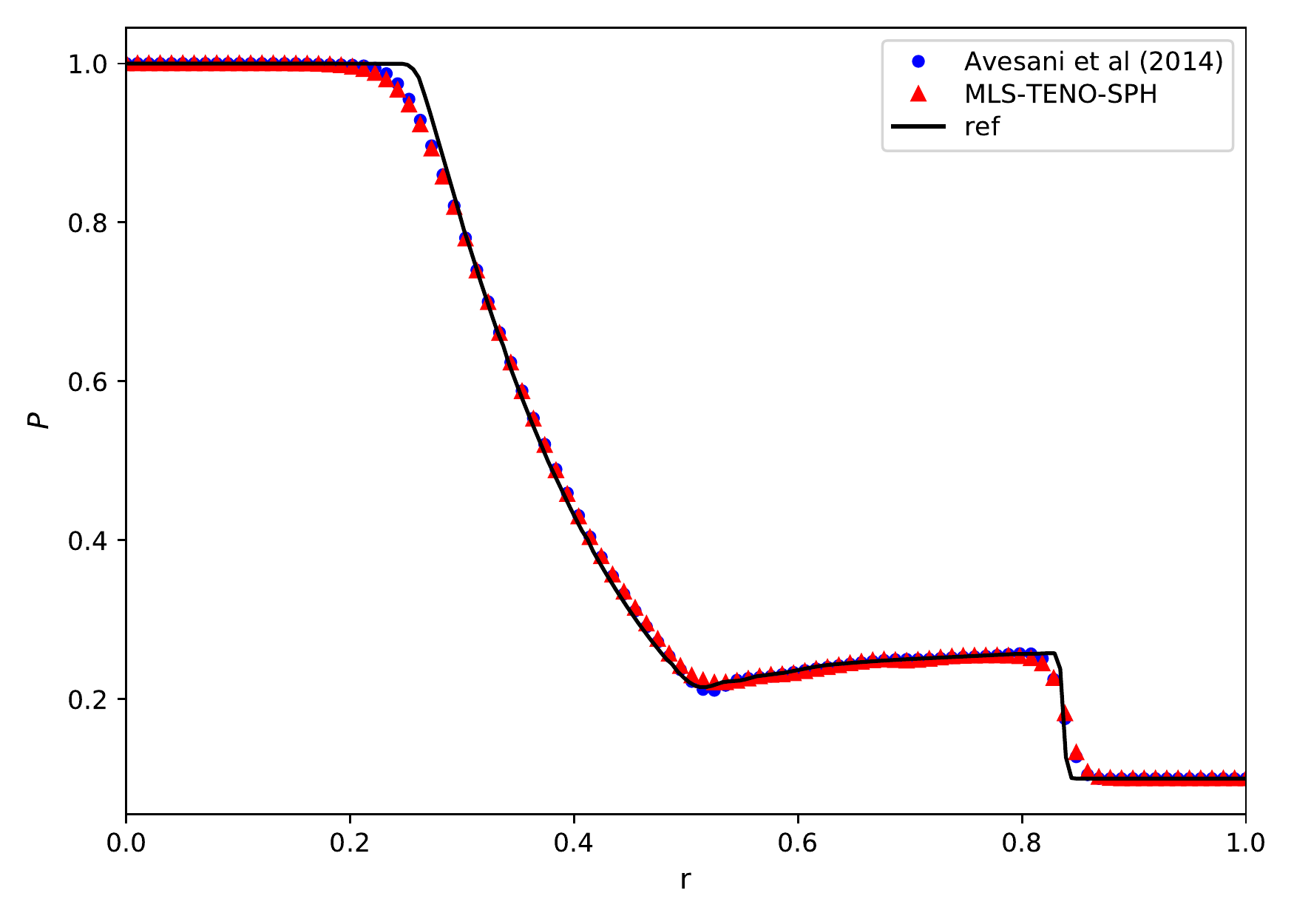}}
		\subfigure[Pressure (Lagragian)]{
			\includegraphics[width=0.45\linewidth]{ 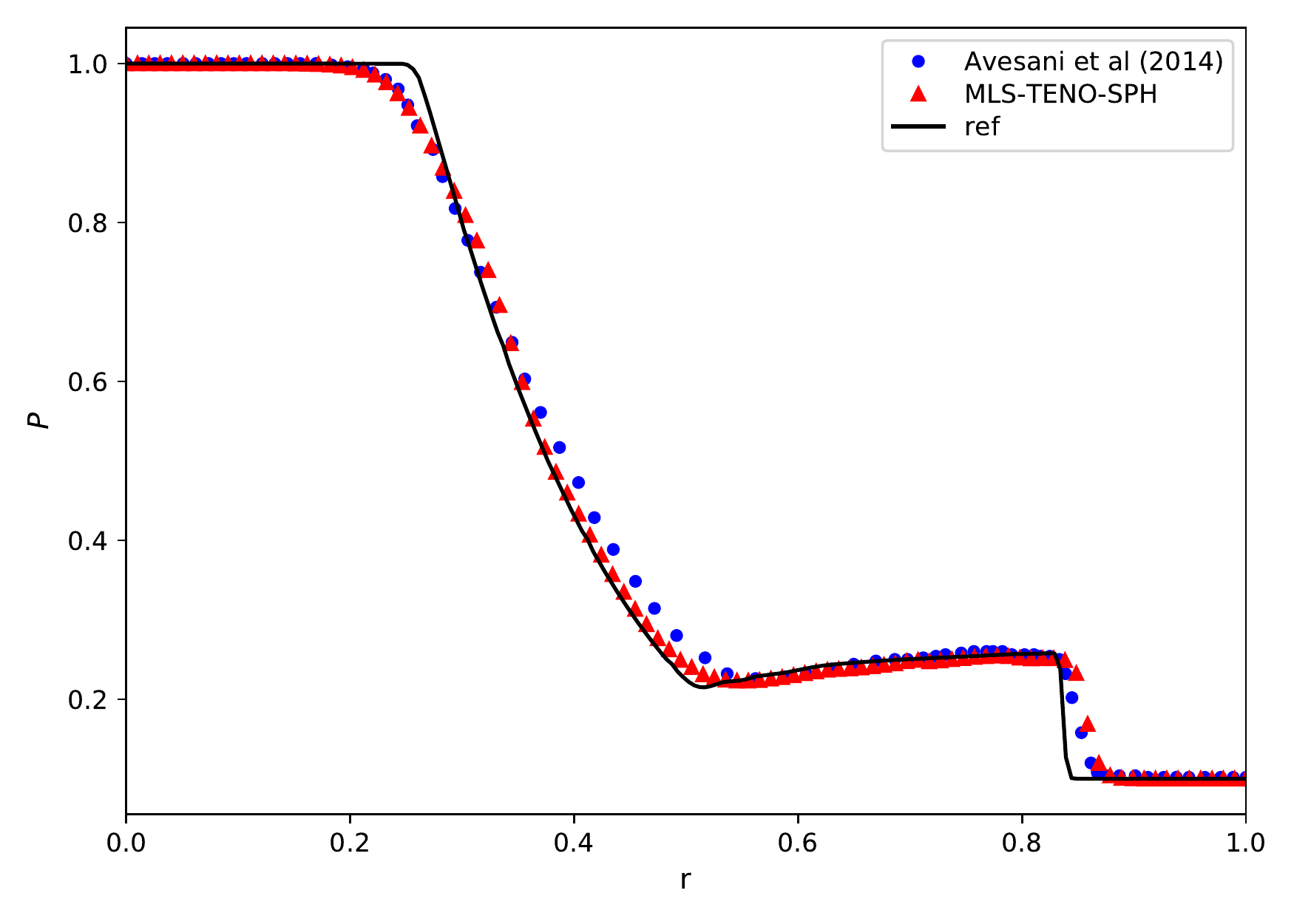}}
		
		\caption{Circular blast wave problem: density and pressure plots along radial direction simulated by the present MLS-TENO-SPH method in the Eulerian and Lagrangian frameworks, and the comparison to the MLS-WENO-SPH method proposed by Avesani et al. \cite{avesani2014new}. }
		\label{bw_plt}
	\end{figure*}

	Fig. \ref{cnt_bw} shows the density and pressure fields calculated with the present MLS-TENO-SPH method in the Eulerian and Lagrangian frameworks. Both simulations show  consistent results without obvious numerical oscillations. This case is also investigated with the method of Avesani et al. \cite{avesani2014new}. For quantitative comparisons, we plot the density and pressure distributions along the radial direction, as shown in Fig. \ref{bw_plt}. In the Eulerian framework, the two methods present similar results. In the  Lagrangian framework, the present method has a better performance at around $ \left|\boldsymbol{r} \right|=0.5 $. Also, a sharper contact discontinuity is captured with both methods at around $ \left|\boldsymbol{r} \right|=0.7 $ in the Lagrangian framework, which is usually due to the denser particle distribution at this location. 
	
	The computational costs of the present MLS-TENO-SPH method and the MLS-WENO-SPH method proposed by Avesani et al. \cite{avesani2014new} in the Eulerian and Lagrangian framework are compared in Table \ref{tb:bw}. It is shown that the computational costs of the present simulation only account for around $ 40\% $ of the MLS-WENO-SPH method by Avesani et al. \cite{avesani2014new}, while the present method shows almost the same results in the Eulerian framework and even better accuracy in the Lagrangian framework. 
	\begin{table}
		\renewcommand{\arraystretch}{1.3}
		\caption{Circular blast wave problem: comparisons of computational time  using  the present MLS-TENO-SPH (O4) method and the MLS-WENO-SPH method proposed by Avesani et al. \cite{avesani2014new} in the Eulerian and Lagrangian frameworks.} 
		\begin{center}
			\begin{tabular}{p{3cm} p{5cm} p{5cm}}
				\toprule [1.2 pt]
				& MLS-WENO-SPH (O4) & MLS-TENO-SPH (O4) \\
				\hline
				Eulerian & 44 min & 13.18 min \\	
				Lagrangian & 51.65 min & 19.93 min \\	
				\bottomrule [1.2 pt]
			\end{tabular}
		\end{center}
		\label{tb:bw}
	\end{table}

	\subsection{Rayleigh-Taylor instability}
	Rayleigh-Taylor instability is usually adopted to examine the stability and dissipation property of a numerical method. Both the Eulerian and the ALE frameworks are employed to simulate this problem. In a domain $\left[  0, 0.25 \right] \times \left[  0, 1 \right]  $, the high-density fluids will penetrate into the low-density fluids under the gravity effect. 
	The initial condition of this case is \cite{xu2005anti}
	\begin{equation}
	(\rho, u, v, p)=\left\{\begin{array}{cl}
	(2,0,-0.025 c \cos (8 \pi x), 1+2 y), & \text { if } 0 \leq y<1 / 2 \\
	(1,0,-0.025 c \cos (8 \pi x), y+3 / 2), & \text { if } 1 / 2 \leq y \leq 1
	\end{array}\right.
	\end{equation}
	and $ c $ is the sound speed $c= \sqrt{\gamma \frac{p}{\rho}} $, where $ \gamma=\frac{5}{3} $. The gravity is $ g=1 $ in the vertical upward direction. The reflective boundary condition is deployed at the left and right boundaries. At the top boundary, $(\rho, u, v, p) $ is set as $(1, 0, 0, 2.5) $ and at the bottom boundary, $(\rho, u, v, p) $ is set as $(2, 0, 0, 1) $. 
	
\begin{figure*}[htbp]
		\centering
		
		\subfigure[]{
			\includegraphics[height=0.65\linewidth]{ 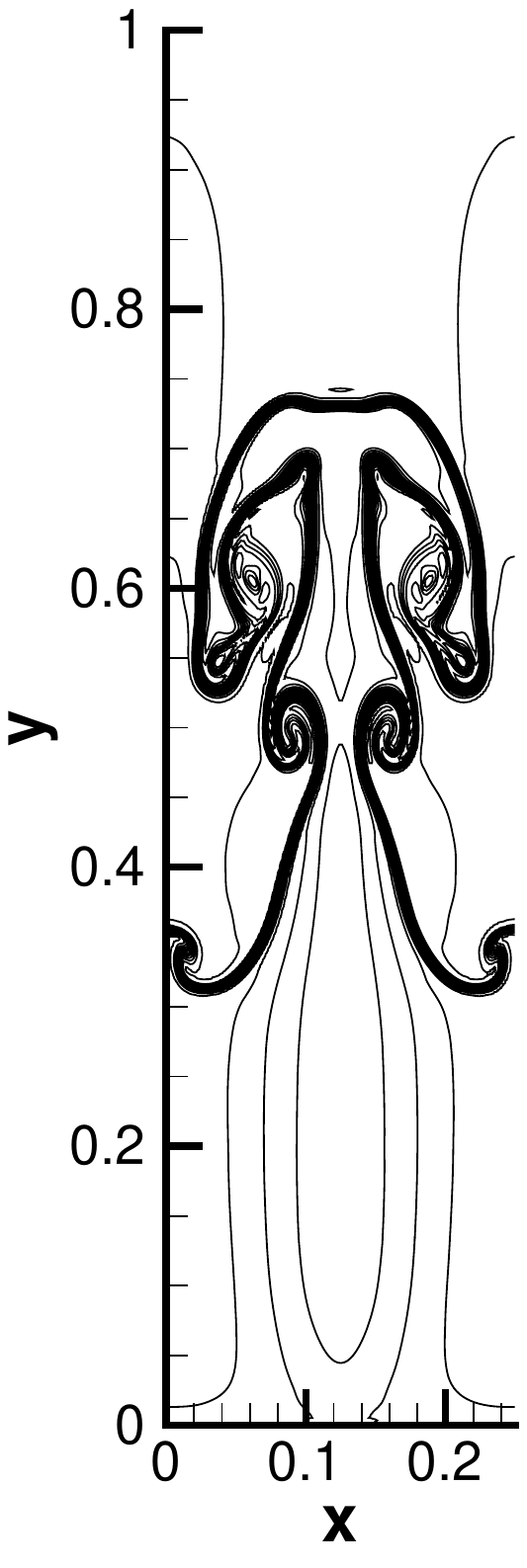}}
		\subfigure[]{
			\includegraphics[height=0.65\linewidth]{ 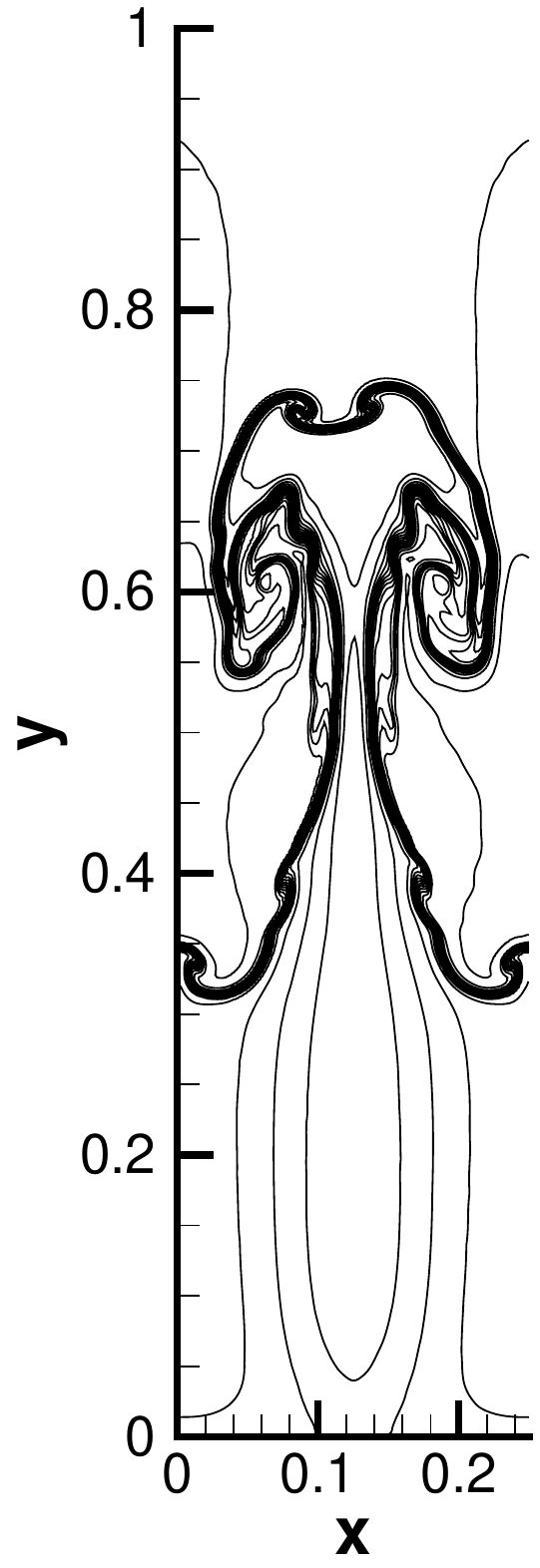}}	
		\subfigure[]{
			\includegraphics[height=0.65\linewidth]{ 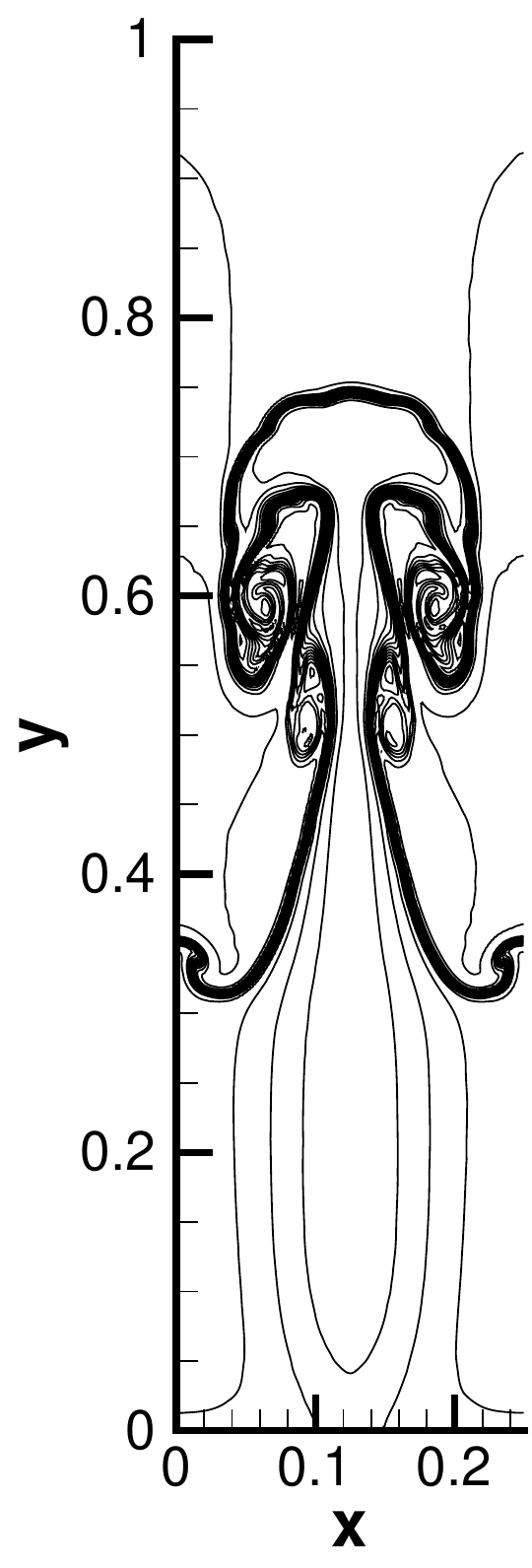}}
		\subfigure[]{
			\includegraphics[height=0.65\linewidth]{ 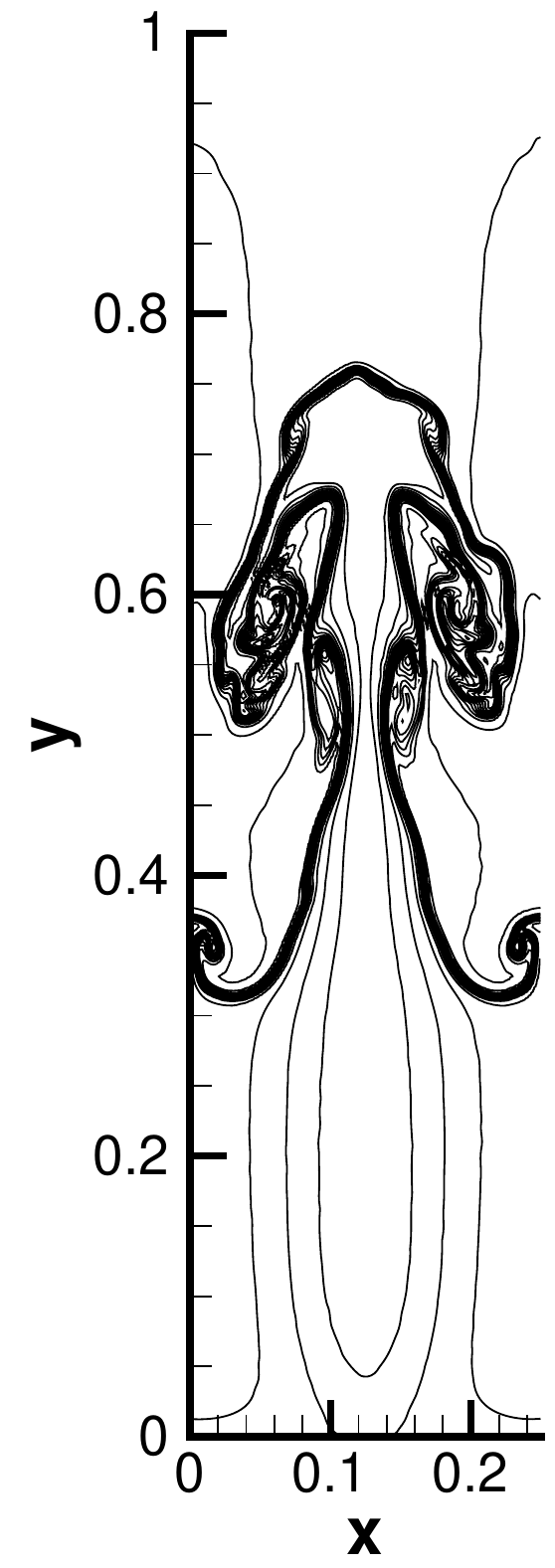}}
		
		\caption{Rayleigh-Taylor instability: density fields simulated with (a) the MLS-WENO-SPH method \cite{avesani2014new}, and the comparisons to the present MLS-TENO-SPH methods of different reconstruction orders, i.e., (b) fourth-order (O4), (c) fifth-order (O5)  and (d) sixth-order (O6)  in the Eulerian framework. This figure is drawn with 15 density contour lines between 0.9 and 2.1. }
		\label{rt}
	\end{figure*}

	Fig. \ref{rt} presents the density fields simulated with the MLS-WENO-SPH method and the comparisons to the present MLS-TENO-SPH methods of different orders in the Eulerian framework. The lattice particle distribution is used, and the resolution is set as $ \Delta x=1/128 $ for all the cases. Comparing Fig. \ref{rt}(a) and (b), the performance of two methods with the same formal order is almost the same, while the present method can reduce computational costs significantly as shown in Table \ref{tb:rt}. As shown in Fig. \ref{rt}(b) and (d),  increasing the reconstruction order of the central large stencil leads to more small-scale structures at the top of interfaces and near the vortex regions, indicating less numerical dissipation. 

\begin{figure}[htbp]
		\centering
		\includegraphics[width=0.8\textwidth]{ 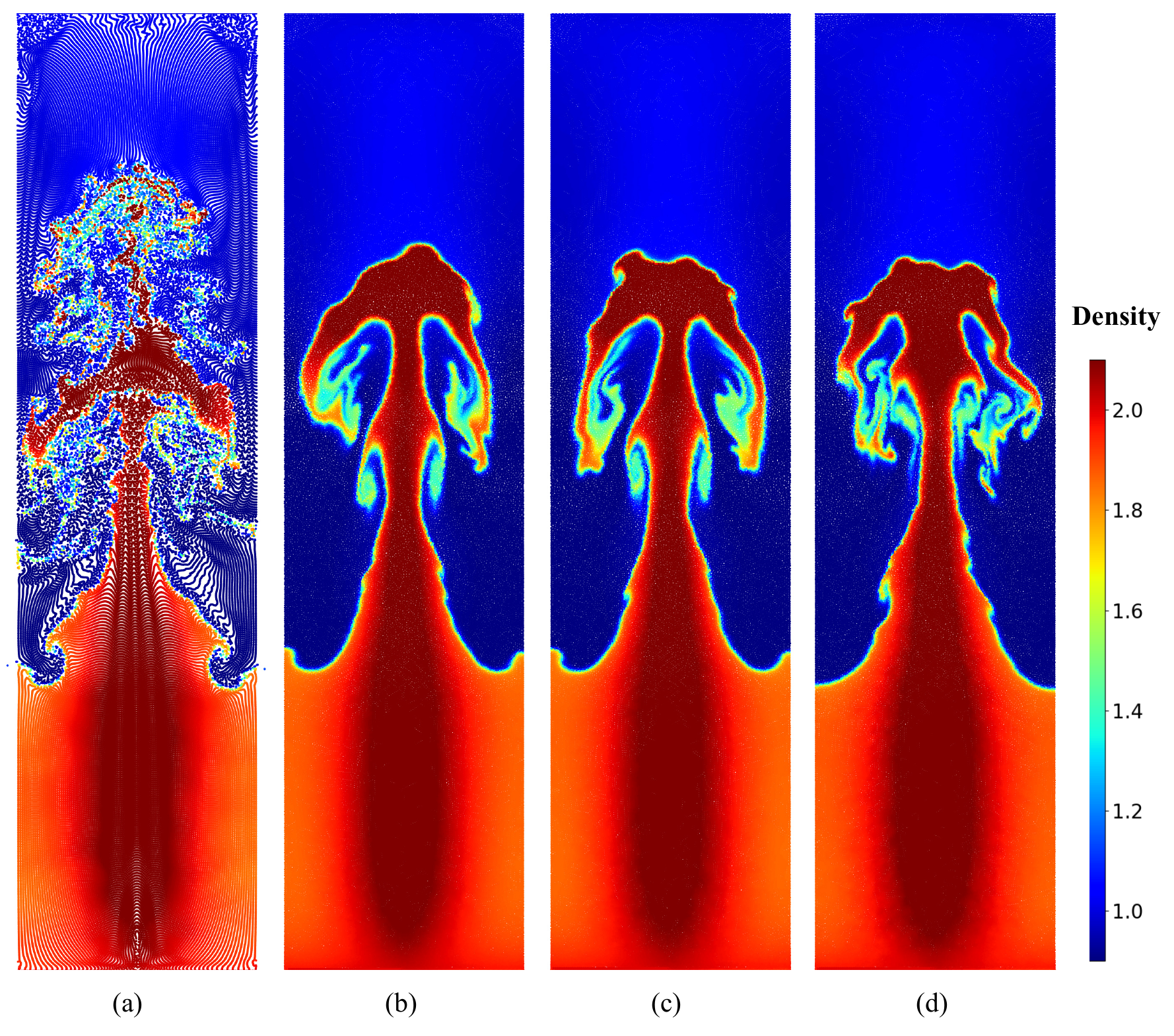}
		\caption{Rayleigh-Taylor instability: results from (a) the MLS-WENO-SPH method \cite{avesani2014new}, and the comparisons to the present MLS-TENO-SPH methods of different reconstruction orders, i.e., (b) fourth-order (O4), (c) fifth-order (O5) and (d) sixth-order (O6) in the ALE framework. }
		\label{rt_ale}
	\end{figure}

	Fig. \ref{rt_ale}(a) shows the results simulated using the MLS-WENO-SPH method based on the fully Lagrangian framework, and a highly disordered particle distribution is observed near the interfaces.
	The ALE framework with the properly designed transport velocity is also employed for the simulation using the present MLS-TENO-SPH method, where the order of the central stencil is fourth. The density field is shown in Fig. \ref{rt_ale}(b), and the computed density distribution is consistent with that in Fig. \ref{rt}(b). Thanks to the  transport velocity deployed in the ALE framework, the particle distribution is rather isotropic, and the small-scale flow structures near the interfaces are also well resolved. In Fig. \ref{rt_ale}(c), a higher fifth-order MLS-TENO-SPH method (O5) results in more fine-scale flow structures at the top of interfaces. The same conclusion applies to the sixth-order MLS-TENO-SPH method (O6), as shown in Fig. \ref{rt_ale}(d). Regarding the efficiency, it is shown that the present framework takes significantly less computational time than MLS-WENO-SPH method as shown in Table \ref{tb:rt}. 

	\begin{table}
		\renewcommand{\arraystretch}{1.3}
		\caption{Rayleigh-Taylor instability: comparisons of the computational time  using  the present MLS-TENO-SPH method and the MLS-WENO-SPH method proposed by Avesani et al. \cite{avesani2014new} with fixed and moving particles.} 
		\begin{center}
			\begin{tabular}{p{3cm} p{5cm} p{5cm}}
				\toprule [1.2 pt]
				& MLS-WENO-SPH (O4) & MLS-TENO-SPH (O4)  \\
				\hline
				Fixed & 512 min (Eulerian) & 210 min (Eulerian) \\	
				Moving & 1145 min (Lagrangian) & 919 min (ALE) \\	
				\bottomrule [1.2 pt]
			\end{tabular}
		\end{center}
		\label{tb:rt}
	\end{table}

	\subsection{Kelvin-Helmholtz instability}
	
	The Kelvin-Helmholtz instability is usually induced by shear stress at the interface, and the interface may develop complex unstable vortical structures \cite{san2015evaluation}\cite{frank1995mhd}. The computational domain is $ \left[-0.5, 0.5 \right]\times \left[-0.5, 0.5 \right] $, and the initial condition is 
	\begin{equation}
	(\rho, u, v, p)=\left\{\begin{array}{cc}
	(2,-0.5, 0.01\sin(2\pi x/L), 2.5), & |y| \leq 0.25, \\
	(1,0.5, 0.01\sin(2\pi x/L), 2.5), & \text { otherwise, }
	\end{array}\right.
	\end{equation} 
	and $ \gamma=1.4 $. The periodic boundary conditions are imposed on all the boundaries in this case. The final simulation time is set as $t = 1$. The resolution is set as $ \Delta x=1/256 $ for the simulations. 
	
\begin{figure}[htbp]
		\centering
		\includegraphics[width=0.95\textwidth]{ 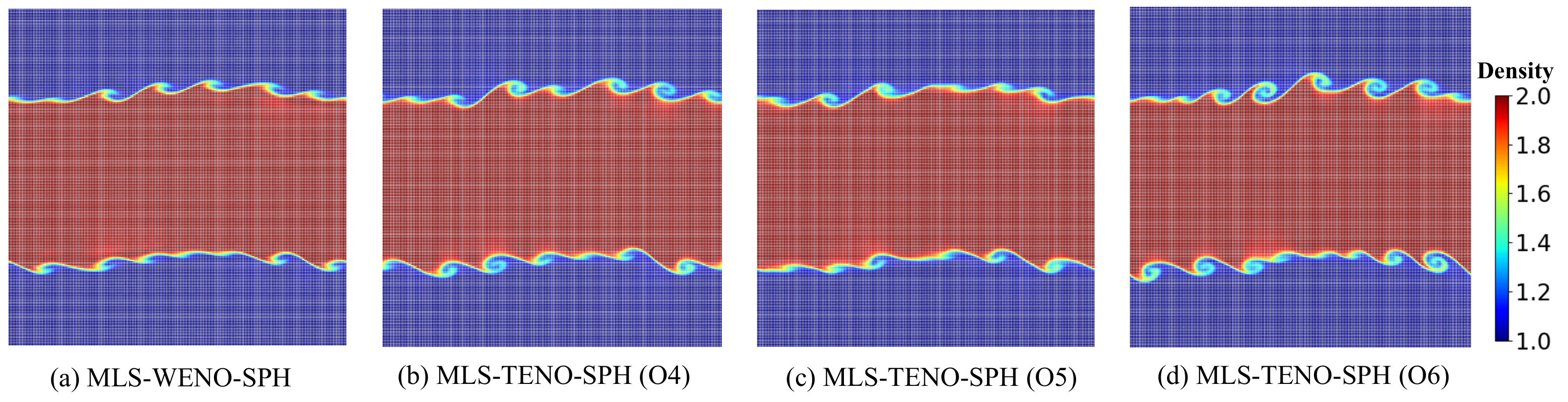}
		\caption{Kelvin-Helmholtz instability: density fields simulated with (a) the MLS-WENO-SPH method \cite{avesani2014new}, and the comparisons to the present MLS-TENO-SPH methods of different reconstruction orders, i.e., (b) fourth-order (O4), (c) fifth-order (O5)  and (d) sixth-order (O6), in the Eulerian framework. }
		\label{kh_e}
	\end{figure}

	Fig. \ref{kh_e} shows the density fields simulated with the MLS-WENO-SPH method \cite{avesani2014new}, and the comparisons to the present MLS-TENO-SPH methods in the Eulerian framework. As shown in Fig. \ref{kh_e}(a) and (b), obviously, more vortical structures are captured by the present method with the same accuracy order. It is further demonstrated that higher-order MLS-TENO-SPH methods predict more fine-scale structures with less numerical dissipation, as shown in Fig. \ref{kh_e}(c) and (d).  The computational cost statistics in Table \ref{tb:kh} show that the present MLS-TENO-SPH method saves significant computational time when compared to the MLS-WENO-SPH method in the Eulerian framework. 
	
\begin{figure}[htbp]
		\centering
		\includegraphics[width=0.95\textwidth]{ 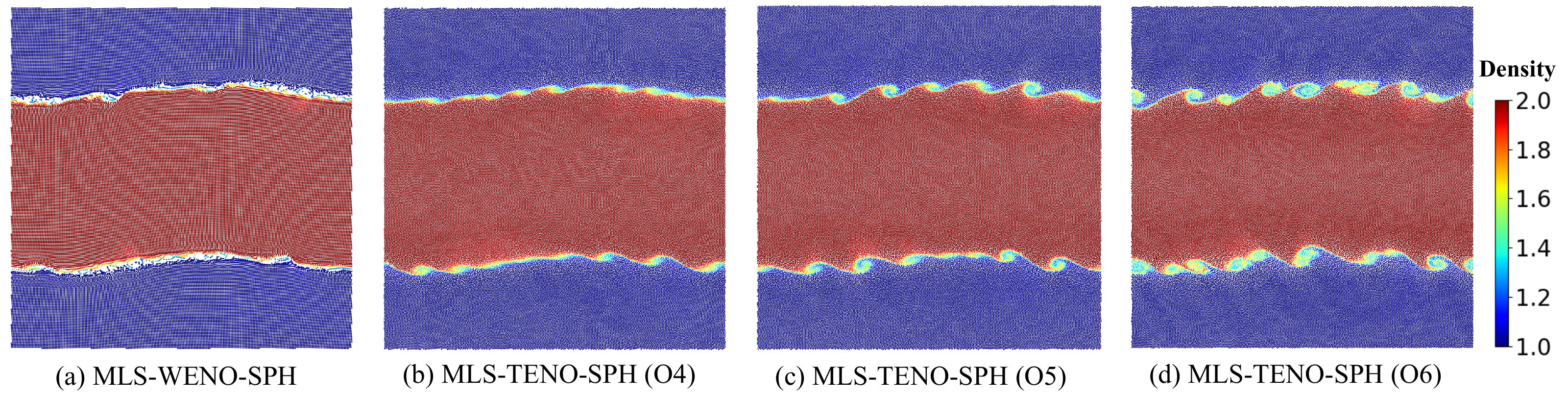}
		\caption{Kelvin-Helmholtz instability: density fields simulated with (a) the MLS-WENO-SPH method \cite{avesani2014new}, and the comparisons to the present MLS-TENO-SPH methods of different reconstruction orders, i.e., (b) fourth-order (O4), (c) fifth-order (O5)  and (d) sixth-order (O6)  in the ALE framework. }
		\label{kh_ale}
	\end{figure}

	Fig. \ref{kh_ale}(a) shows the density fields simulated with the conventional MLS-WENO-SPH method \cite{avesani2014new} in the fully Lagrangian framework. While a disordered particle distribution is shown near the interfaces, almost no small-scale vortical structures are captured by MLS-WENO-SPH. By contrast, using the present MLS-TENO-SPH method in the ALE framework, more vortical structures present, and the particle distribution is also much more isotropic, as shown in Fig. \ref{kh_ale}(b). Further enhancing the reconstruction order of present MLS-WENO-SPH methods, as shown in Fig. \ref{kh_ale}(c) and (d), the prediction of interfacial instability is significantly improved. The computational cost statistics in Table \ref{tb:kh} shows that the present MLS-TENO-SPH method is also more efficient.

	\begin{table}
		\renewcommand{\arraystretch}{1.3}
		\caption{Kelvin-Helmholtz instability: comparisons of computational time  using  the present MLS-TENO-SPH method and the MLS-WENO-SPH method proposed by Avesani et al. \cite{avesani2014new} with fixed and moving particles.} 
		\begin{center}
			\begin{tabular}{p{3cm} p{5cm} p{5cm}}
				\toprule [1.2 pt]
				& MLS-WENO-SPH (O4) & MLS-TENO-SPH (O4)  \\
				\hline
				Fixed & 290 min (Eulerian) & 208 min (Eulerian)  \\	
				Moving & 257 min (Lagrangian) &  228 min (ALE) \\	
				\bottomrule [1.2 pt]
			\end{tabular}
		\end{center}
		\label{tb:kh}
	\end{table}

	\subsection{Multi-material triple-point shock problem}
\begin{figure}[htbp]
		\centering
		\includegraphics[width=0.5\textwidth]{ 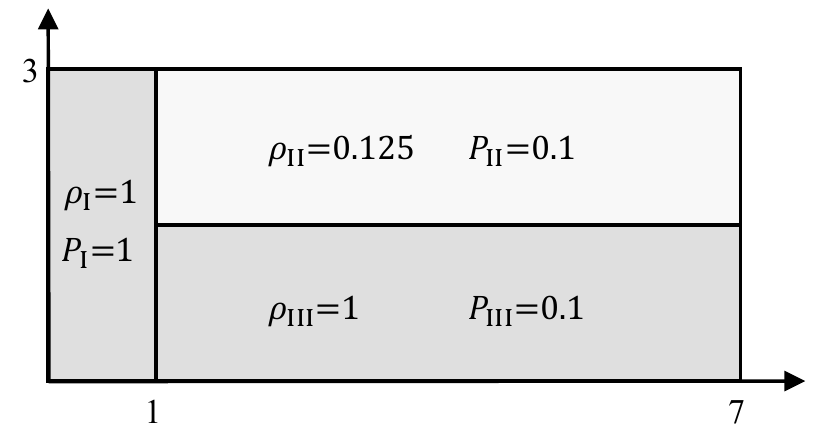}
		\caption{The schematic of a 2D multi-material triple-point shock problem, where the initial fluid velocity is set as zero.  }
		\label{schm_trp}
	\end{figure}

	The 2D multi-material triple-point shock problem is considered in this section. As shown in Fig. \ref{schm_trp}, three materials are initially partitioned into three regions in the rectangle with the length 7 and height 3. Following the setup of \cite{zeng2014frame}, in the left region, the field values are $ (\rho_{\rm\uppercase\expandafter{\romannumeral1}}=1, P_{\rm\uppercase\expandafter{\romannumeral1}}=1 ) $; in the upper right region, the field values are $ (\rho_{\rm\uppercase\expandafter{\romannumeral2}}=0.125, P_{\rm\uppercase\expandafter{\romannumeral2}}=0.1 ) $, and in the bottom right region, the field values are $ (\rho_{\rm\uppercase\expandafter{\romannumeral3}}=1, P_{\rm\uppercase\expandafter{\romannumeral3}}=0.1 ) $. The adiabatic index $ \gamma $ is chosen as $ 1.4 $. The high-pressure material $\rm I $ penetrates into the upper right area due to the low density in the region $\rm {II} $, and then a shear flow is generated at the intersection point of the three materials. The reflective boundary condition is deployed for all the boundaries. In this case, the resolution is set as $ \Delta x=3/384 $. The Eulerian framework is employed for the simulation in this case.

\begin{figure*}[htbp]
		\centering
		
		\subfigure[MLS-WENO-SPH (O4)]{
			\includegraphics[width=0.48\linewidth]{ 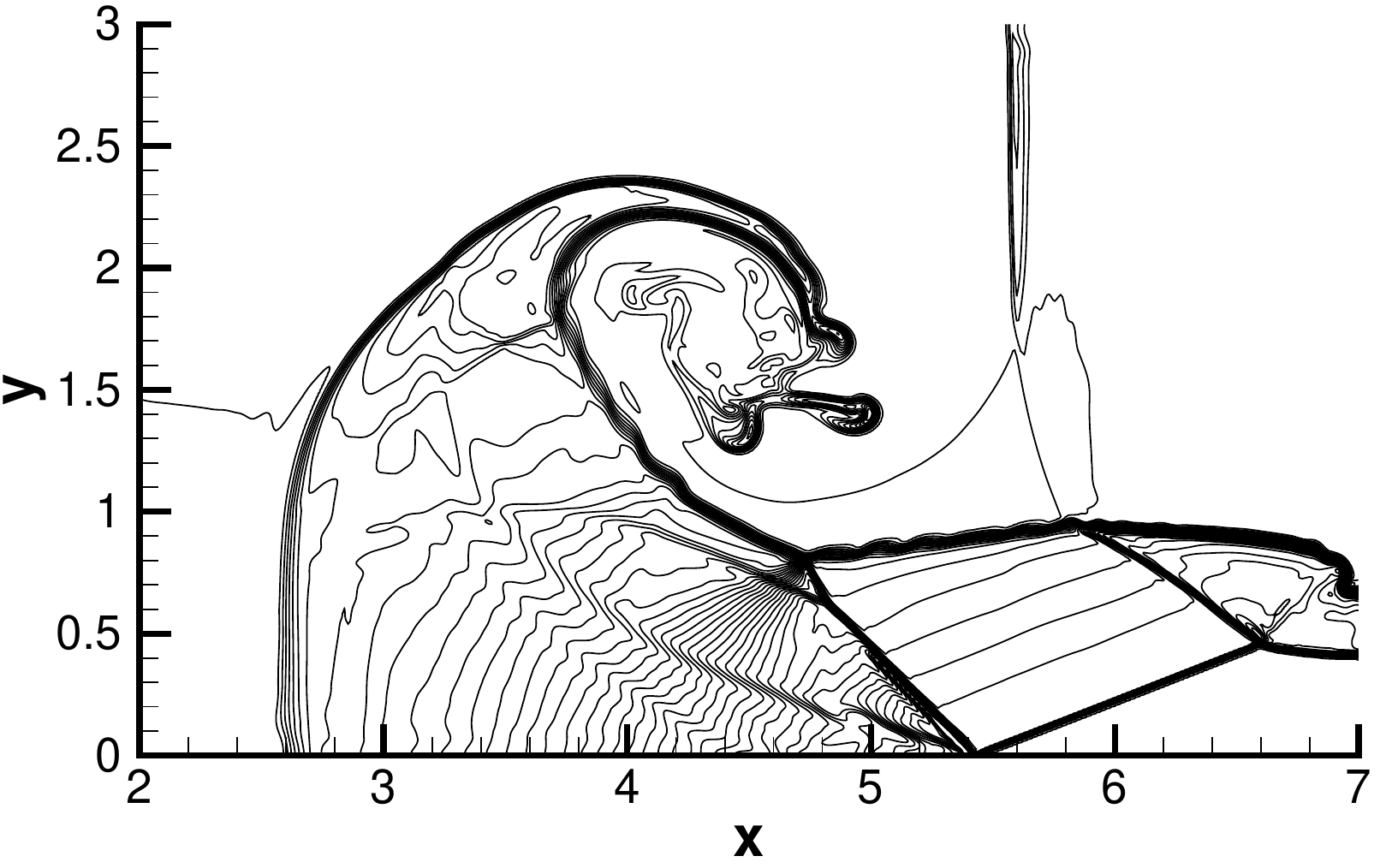}}
		\subfigure[MLS-TENO-SPH (O4) ]{
			\includegraphics[width=0.48\linewidth]{ 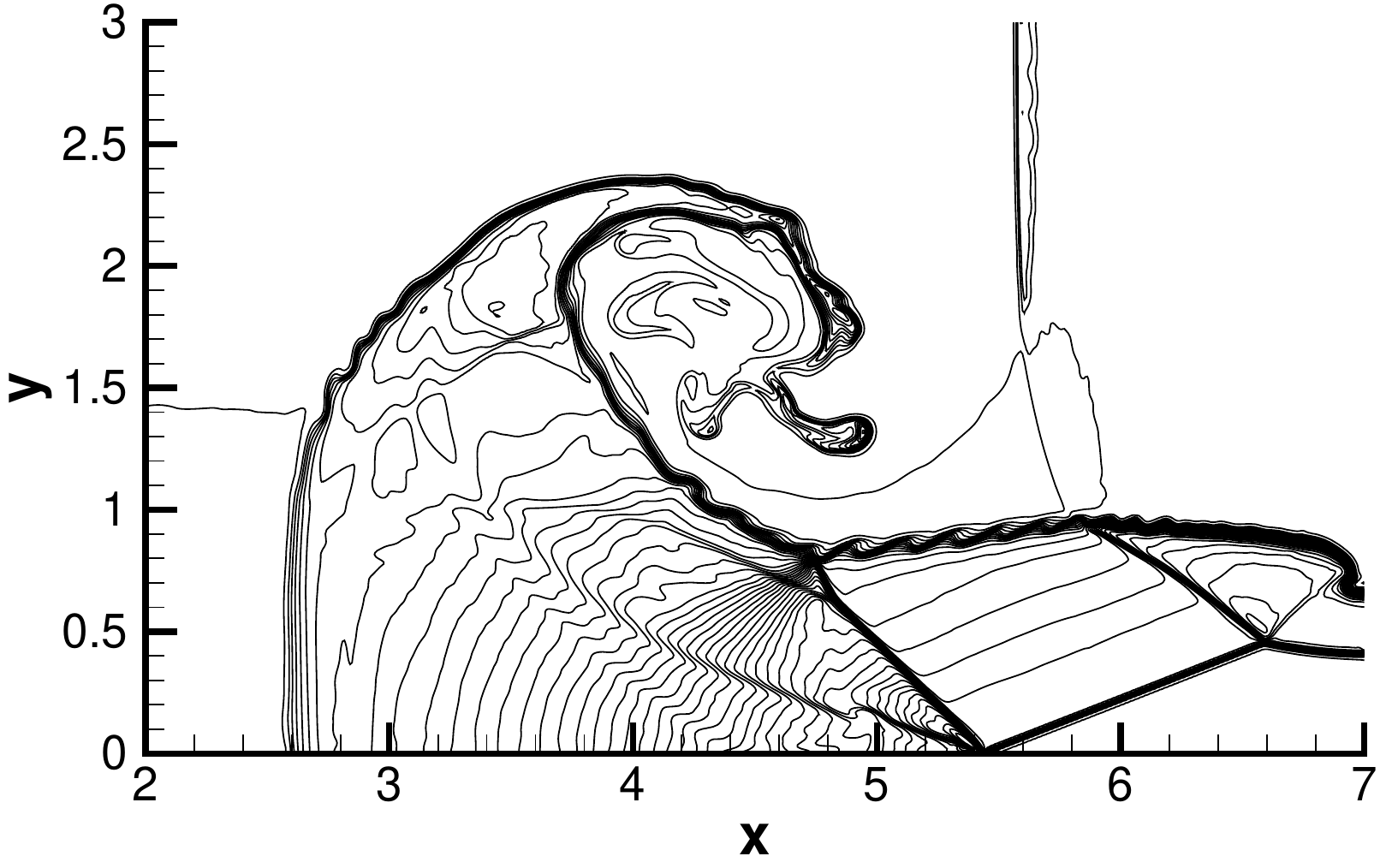}}
		\subfigure[MLS-TENO-SPH (O5) ]{
			\includegraphics[width=0.48\linewidth]{ 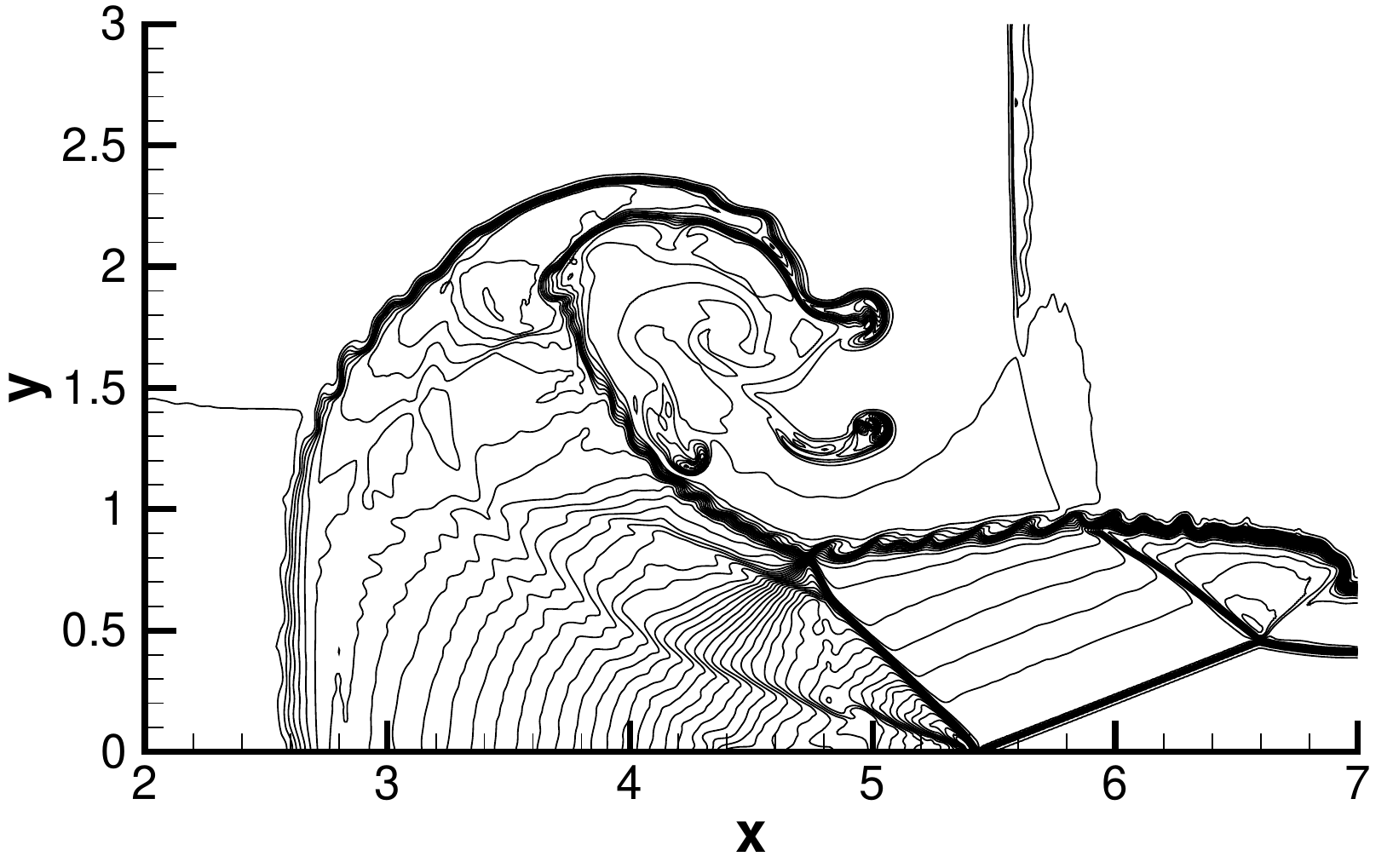}}	
		\subfigure[MLS-TENO-SPH (O5), disorder]{
			\includegraphics[width=0.48\linewidth]{ 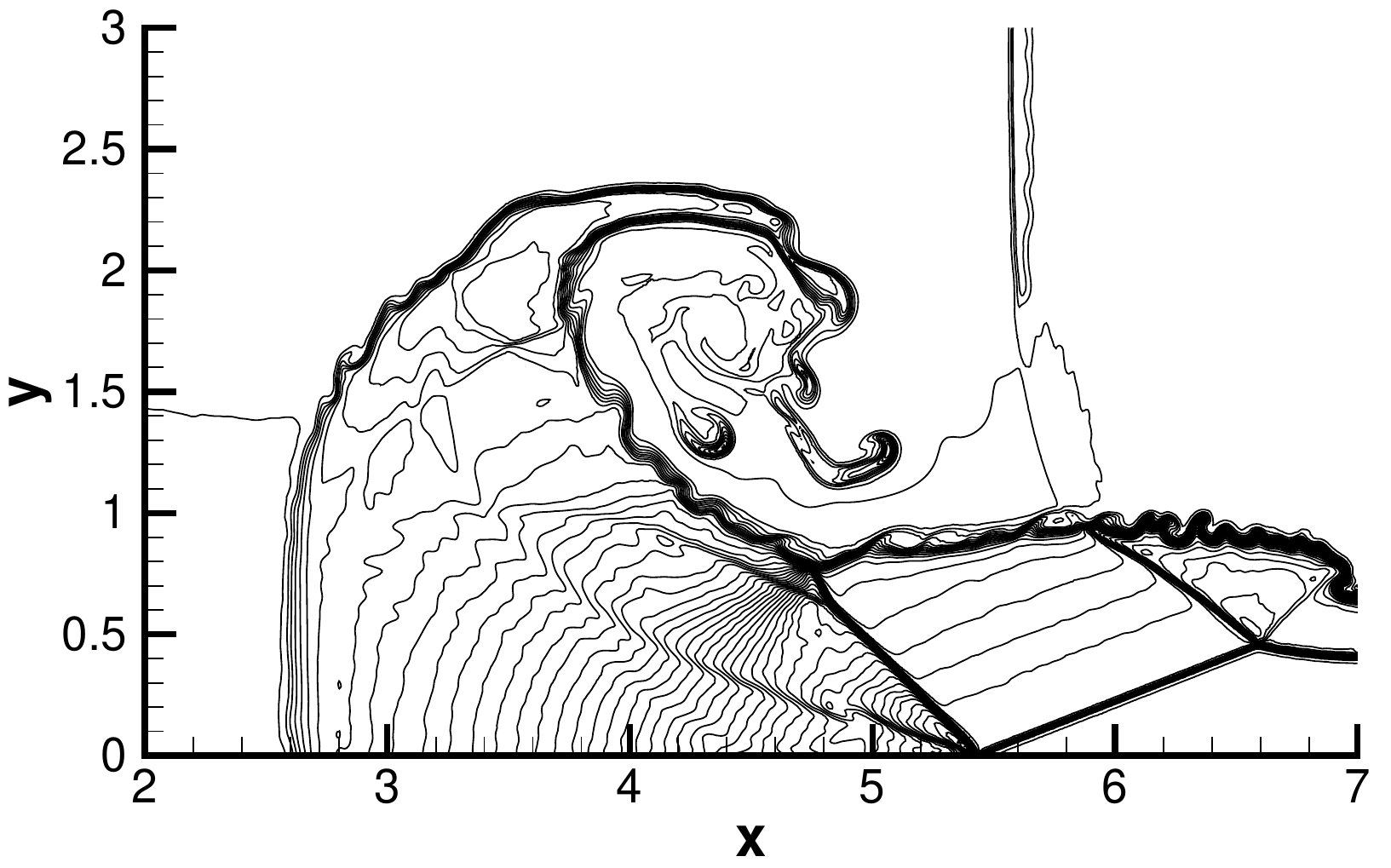}}	  
		
		\caption{Multi-material triple-point shock problem: density fields simulated with (a) the MLS-WENO-SPH method \cite{avesani2014new}, and the comparisons to the present MLS-TENO-SPH methods of different particle configurations and different reconstruction orders, i.e., latticed particle distributions with (b) fourth-order (O4) and (c) fifth-order (O5), and (d) disordered particle distributions with fifth-order (O5) in the Eulerian framework. This figure is drawn with 50 contour lines between 0.125 and 5. }
		\label{tp_comp}
	\end{figure*}

	Fig. \ref{tp_comp} displays the density fields simulated with the MLS-WENO-SPH method \cite{avesani2014new}, and the comparisons to the present MLS-TENO-SPH methods in the Eulerian framework. Using the same particle distribution, as shown in Fig. \ref{tp_comp} (a) and (b), the results simulated by the present MLS-TENO-SPH  method shows more small-scale vortical structures than that simulated by the MLS-WENO-SPH method, indicating the low-dissipation property of MLS-TENO-SPH. In Fig. \ref{tp_comp}(a), even some numerical oscillations exist in the vicinity of  discontinuities using the MLS-WENO-SPH method, and small-scale flow structures are smeared significantly. The result is further enhanced when the reconstruction order is increased for MLS-TENO-SPH, as seen in Fig. \ref{tp_comp}(c).
	In Fig. \ref{tp_comp}(d), a disordered particle distribution is deployed for the simulation. The particles are randomly disordered away from a lattice particle distribution with the variance of $0.4\Delta x$. The simulation result is consistent with that of the latticed particle distribution. Comparing Fig. \ref{tp_comp}(d) with (a), it shows that the present MLS-TENO-SPH method has a better performance than the MLS-WENO-SPH method even with a disordered particle distribution.
	The computational time statistics in Table \ref{tb:trp} show a significant decrease of 56\% with the present MLS-TENO-SPH method when compared to the conventional MLS-WENO-SPH method, which further demonstrates the all-round performance advantage of the present method.

	\begin{table}
		\renewcommand{\arraystretch}{1.3}
		\caption{Multi-material triple-point shock problem: comparisons of computational time  using  the present MLS-TENO-SPH method and the MLS-WENO-SPH method proposed by Avesani et al. \cite{avesani2014new} with fixed particles.} 
		\begin{center}
			\begin{tabular}{p{3cm} p{5cm} p{5cm}}
				\toprule [1.2 pt]
				& MLS-WENO-SPH (O4) & MLS-TENO-SPH (O4)  \\
				\hline
				Fixed & 2426 min  &  1074 min	 \\	
				\bottomrule [1.2 pt]
			\end{tabular}
		\end{center}
		\label{tb:trp}
	\end{table}

	\subsection{Double Mach reflection problem}
	The double Mach reflection problem \cite{woodward1984numerical} is a widely-used benchmark case to assess the shock-capturing schemes in the simulation of compressible flows. A shock wave of Mach number 10 moves rightward with an inclination angle of $ 60^\circ $, and the initial condition is given as
	\begin{equation}
	(\rho, u, v, p)= \begin{cases}(1.4,0,0,1), & \text { if } y<1.732(x-0.1667), \\ (8,7.145,-4.125,116.8333), & \text { otherwise. }\end{cases}
	\end{equation}
	On the top boundary, an exact shock wave  of Mach 10 is prescribed, and on the left inlet and right outlet, the field values are set as equal to post-shock wave and pre-shock wave values, respectively. On the bottom boundary, a reflective boundary condition is implemented  from $ x=0.1667 $ to $ x=4 $, and the post-shock wave values are set from $ x=0 $ to $ x=0.1667 $. In this case, the particle resolution is set as $ \Delta x=1/256 $. The Eulerian framework is employed for all the simulations. 
	
\begin{figure*}[htbp]
		\centering
		
		\subfigure[MLS-WENO-SPH (O4)]{
			\includegraphics[width=0.48\linewidth]{ 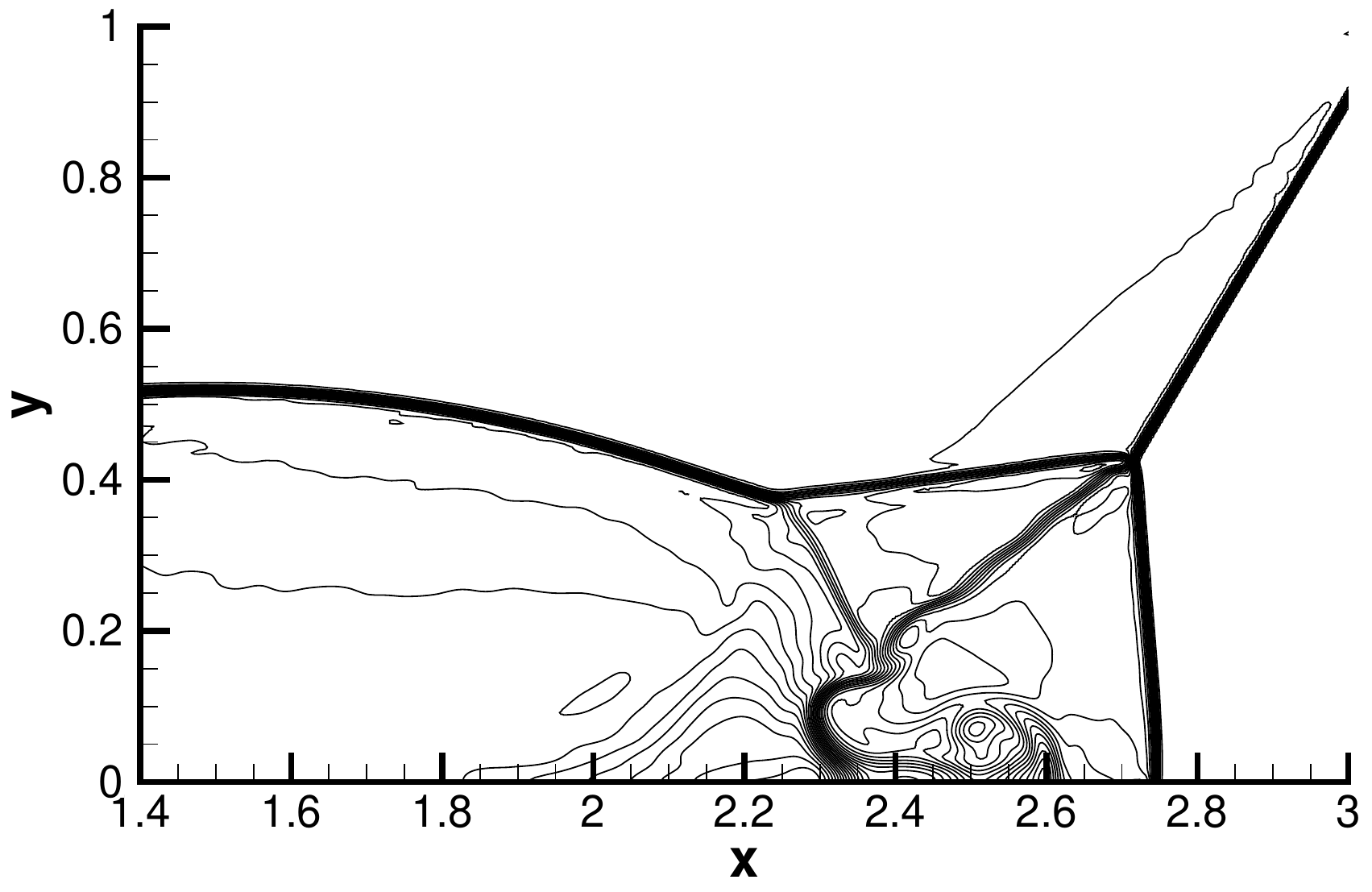}}
		\subfigure[MLS-WENO-SPH (O4), disorder]{
			\includegraphics[width=0.48\linewidth]{ 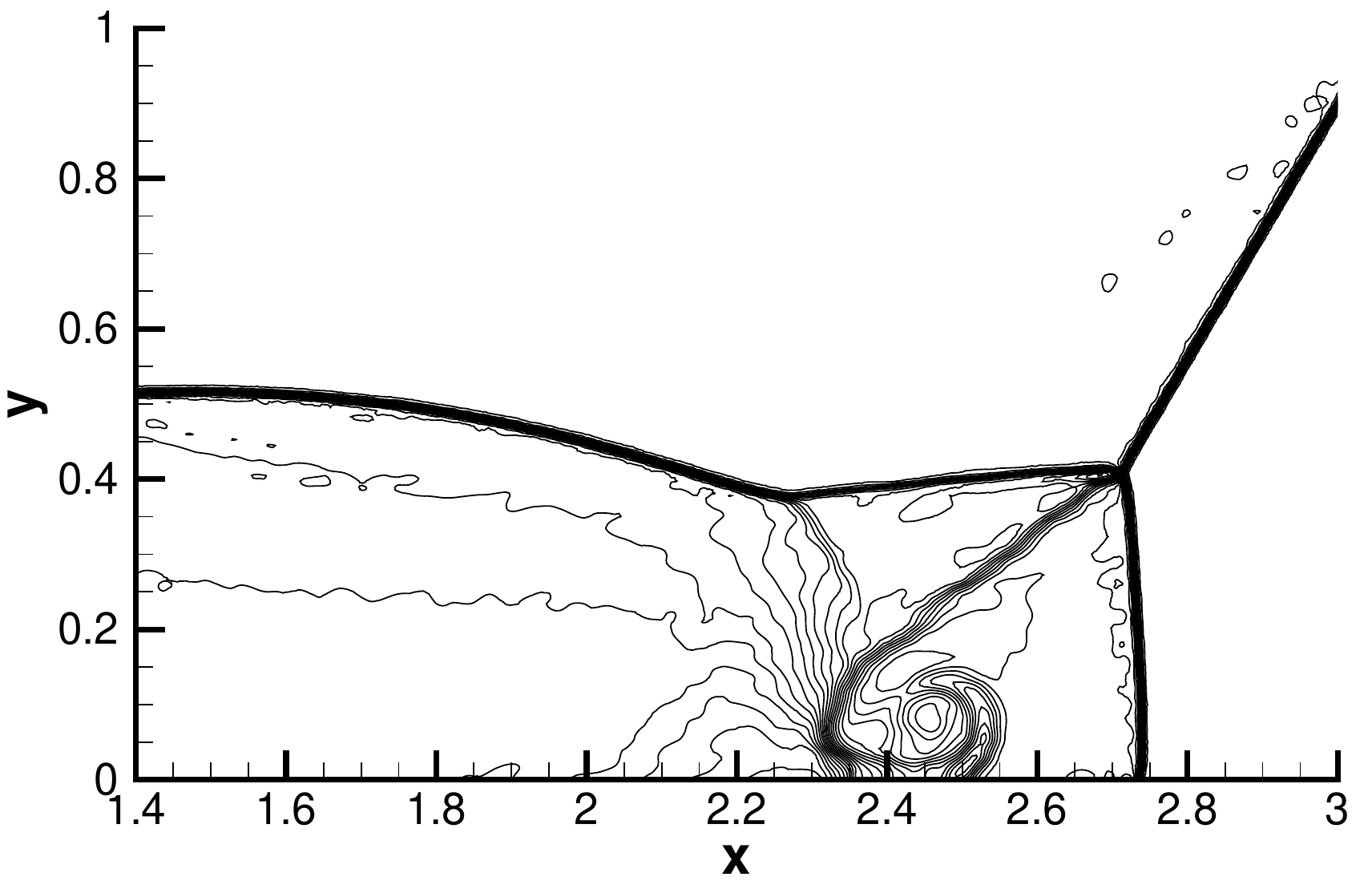}}
		\subfigure[MLS-TENO-SPH (O4)]{
			\includegraphics[width=0.48\linewidth]{ 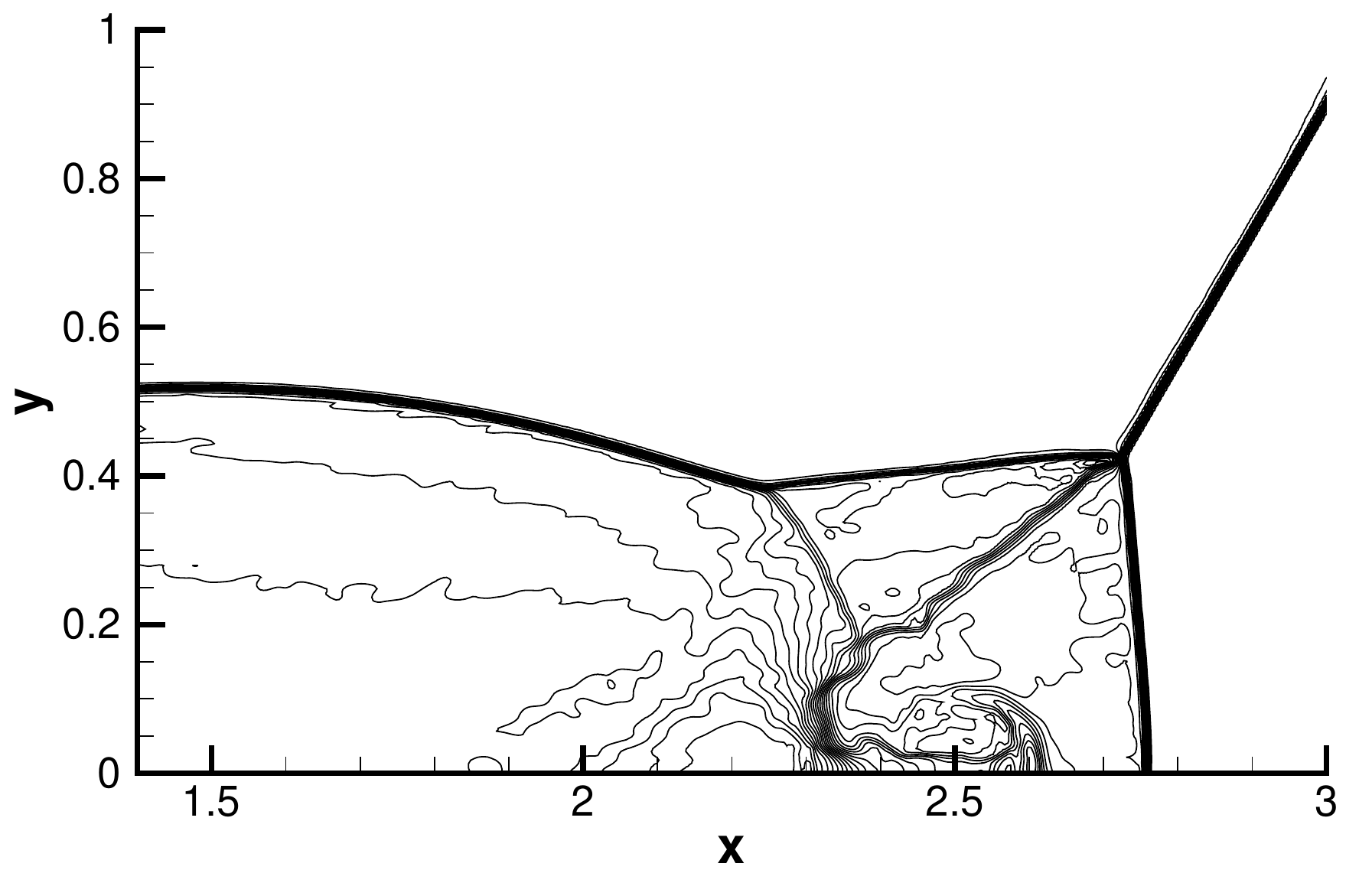}}	
		\subfigure[MLS-TENO-SPH (O4), disorder]{
			\includegraphics[width=0.48\linewidth]{ 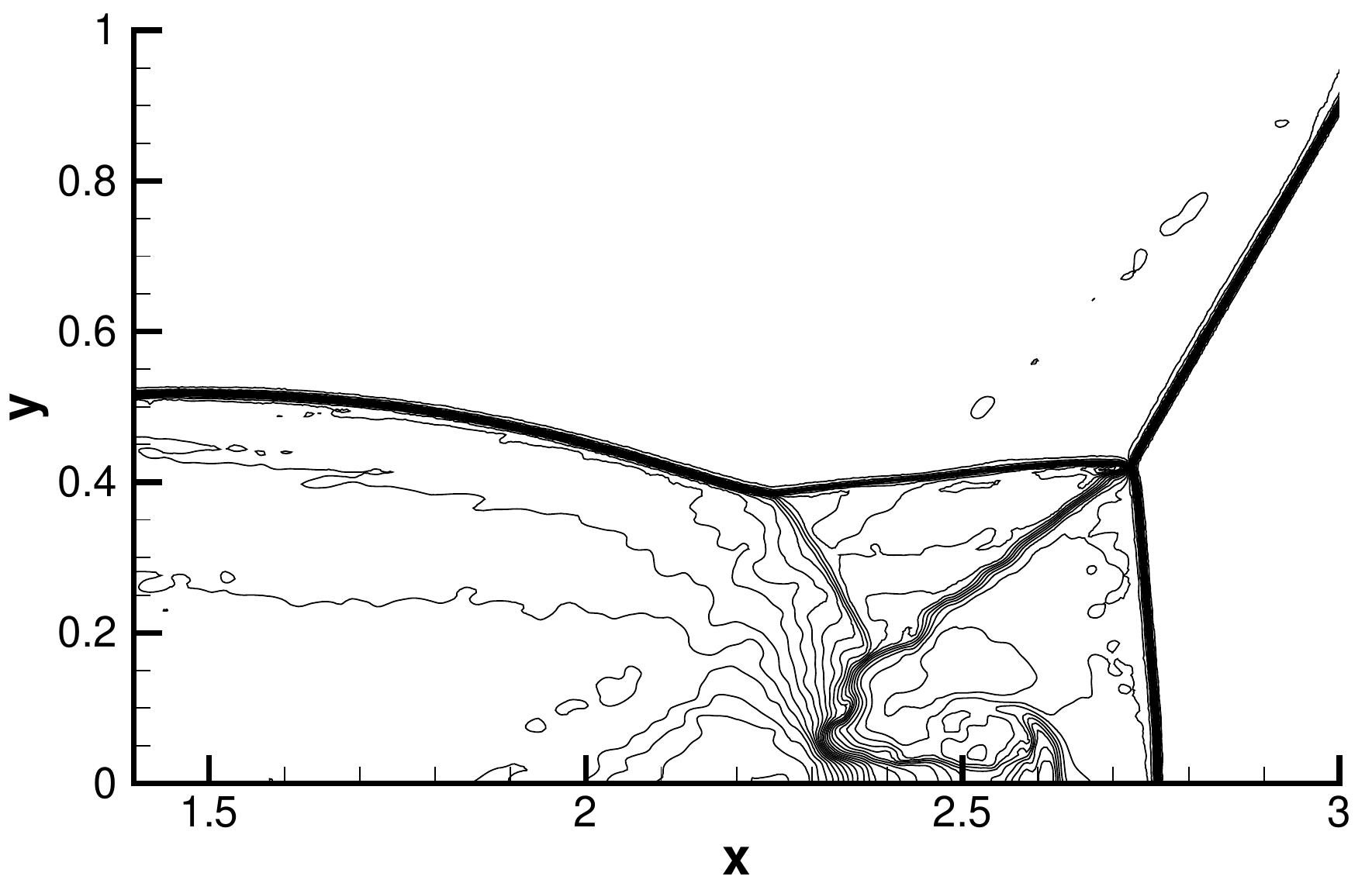}}	   
		\subfigure[MLS-TENO-SPH (O5)]{
			\includegraphics[width=0.48\linewidth]{ 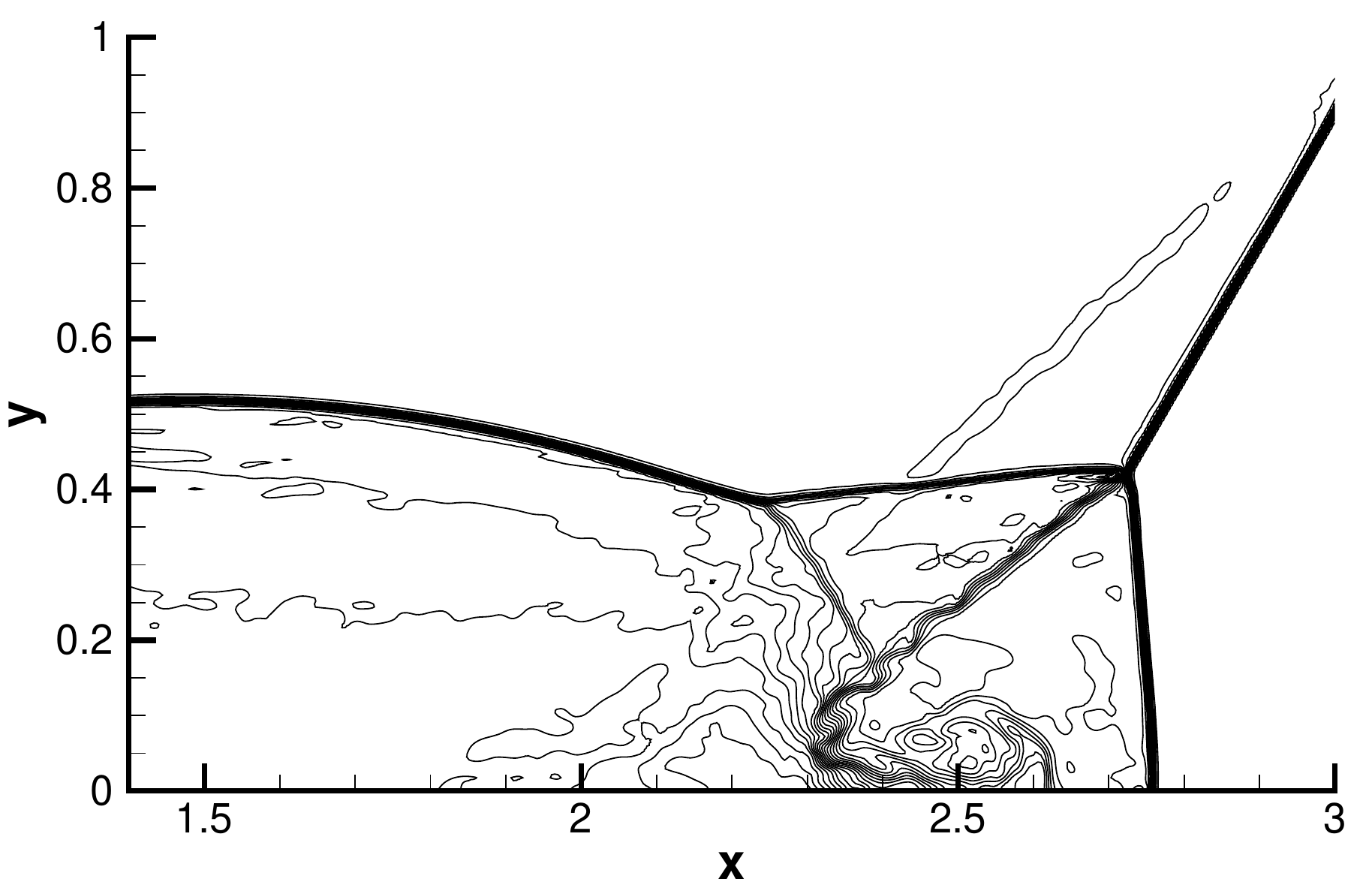}}
		\subfigure[MLS-TENO-SPH (O6) ]{
			\includegraphics[width=0.48\linewidth]{ 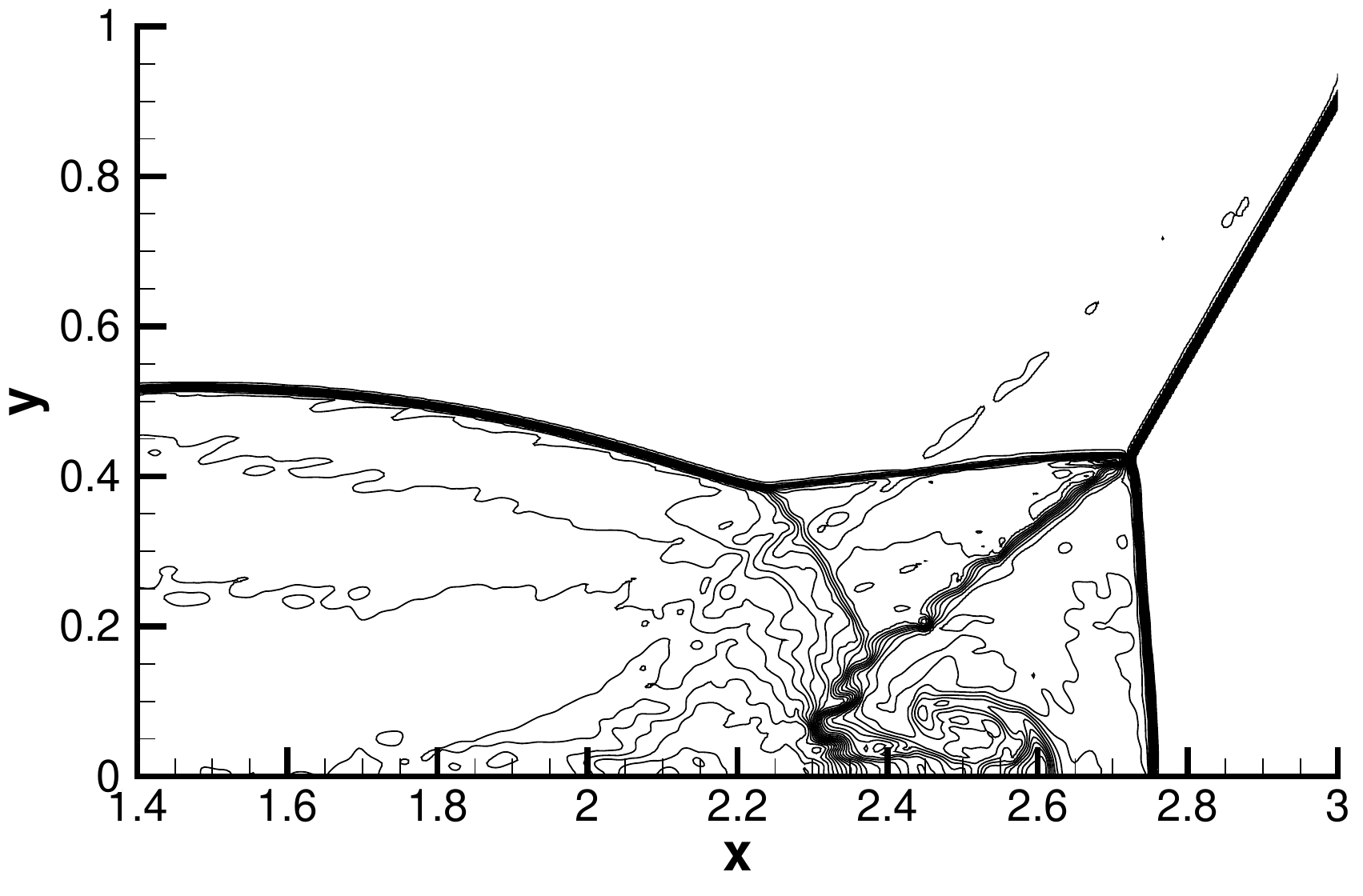}}
		
		\caption{Double Mach reflection problem: density fields simulated by the MLS-WENO-SPH method \cite{avesani2014new} with (a) latticed and (b) disordered particle distributions, MLS-TENO-SPH method with fourth-order reconstruction (O4) using (c) latticed and (d) disordered particle distributions, MLS-TENO-SPH method with (e) fifth-order (O5) and (f) sixth-order (O6) reconstruction using the latticed particle distribution.
        This figure is drawn with 43 contour lines between 1.887 and 20.9.}
		\label{dm}
	\end{figure*}

	Fig. \ref{dm} displays the density fields simulated by the MLS-WENO-SPH and the present MLS-TENO-SPH methods with different reconstruction orders and particle distributions in the Eulerian framework. Comparing  Fig. \ref{dm}(a) and Fig. \ref{dm}(c), it is shown that the present MLS-TENO-SPH method with the fourth-order reconstruction predicts much sharper shock waves and more fine-scale vortical structures, indicating less numerical dissipation than the MLS-WENO-SPH method. When the fifth- and sixth-order MLS-TENO-SPH methods are employed for the simulation, as shown in Fig. \ref{dm}(e) and (f), the prediction is further enhanced. Moreover, as revealed in Fig. \ref{dm}(c), Fig. \ref{dm}(e) and Fig. \ref{dm}(f), no obvious numerical oscillations are observed in the vicinity of discontinuities, and also no carbuncle effects reported in \cite{tsoutsanis2019stencil} take place near the shock waves. 
 
    The results simulated using disordered particle distribution are also shown in Fig. \ref{dm}(b) and Fig. \ref{dm}(d), where the particles are randomly disordered away from a latticed particle distribution with the variance of $0.4\Delta x$. The results simulated by the MLS-WENO-SPH method, as shown in Fig. \ref{dm}(b), have a visible deviation from the other results, which demonstrate that the MLS-WENO-SPH method leads to obvious accuracy degeneration using disorder particle distribution. In contrast, the present MLS-TENO-SPH method can still produce very reliable results using different reconstruction orders, despite a little worse than those simulated using latticed particle distribution. 
	Table \ref{tb:dm} illustrates a substantial decrease of 66\% in the computational costs with the present MLS-TENO-SPH method when compared to the conventional MLS-WENO-SPH method.

	\begin{table}
		\renewcommand{\arraystretch}{1.3}
		\caption{Double Mach reflection problem: comparisons of computational time  using  the present MLS-TENO-SPH method and the MLS-WENO-SPH method proposed by Avesani et al. \cite{avesani2014new} with fixed particles.} 
		\begin{center}
			\begin{tabular}{p{3cm} p{5cm} p{5cm}}
				\toprule [1.2 pt]
				& MLS-WENO-SPH (O4) & MLS-TENO-SPH (O4)  \\
				\hline
				Fixed & 1308 min & 447 min  \\	
				\bottomrule [1.2 pt]
			\end{tabular}
		\end{center}
		\label{tb:dm}
	\end{table}

	\section{Concluding remarks}
	In this study, a new hybrid high-order SPH framework (MLS-TENO-SPH) for compressible flows with discontinuities is developed. This framework can achieve genuine high-order accuracy in smooth regions and also capture discontinuities sharply in non-smooth regions. The framework can be either fully Lagrangian, Eulerian or ALE with the isotropic particle distribution. In the proposed framework, the computational domain is divided into smooth regions and non-smooth regions, and these two regions are determined by a strong scale separation procedure in TENO schemes. In smooth regions, the MLS approximation is used for high-order derivative operator, which restores the genuine high-order reconstruction; in non-smooth regions, the new low-dissipation TENO scheme combining with Vila's SPH framework and several new improvements will be deployed to capture discontinuities and high-wavenumber under-resolved flow scales. The present MLS-TENO-SPH method is validated with a set of challenging cases in the Eulerian, Lagrangian or ALE simulation framework. It is worth noting that some of these cases have never been investigated using high-order SPH methods. Numerical results indicate that the present MLS-TENO-SPH methods of various reconstruction orders are highly efficient, low-dissipation and stable for capturing the shock and contact discontinuities. For most cases considered, the performance improvement is significant when compared to the established MLS-WENO-SPH method.
	
	This study serves as an exploration to develop new high-order SPH methods, and may not be perfect in all respects. \textcolor{black}{Forthcoming work may include the further methodology improvements based on this framework and the deployment of this method to more complicated flows, e.g., compressible multiphase problems with free surfaces. Note that the exact conservation property is not satisfied in the present framework and is also worthy of further investigations. }

	\section*{Acknowledgment}
	
	This work was supported by National Key R\&D Program of China (No. 2022YFA1004500), the Research Grants Council (RGC) of the Government of Hong Kong Special Administrative Region (HKSAR) with RGC/GRF Project (No. 16206321) and RGC/ECS Project (No. 26200222), the fund from Shenzhen Municipal Central Government Guides Local Science and Technology Development Special Funds Funded Projects (No. 2021Szvup138), and the fund from the Project of Hetao Shenzhen-Hong Kong Science and Technology Innovation Cooperation Zone (No. HZQB-KCZYB-2020083).

	
	
	\bibliographystyle{elsarticle-num}
	\scriptsize
	\setlength{\bibsep}{0.5ex}

	\bibliography{bib.bib}
	
\end{document}